\documentclass[a4paper,prd,twocolumn,amsmath,amssymb,aps,preprintnumbers,showpacs]{revtex4}

\usepackage{graphicx}
\usepackage{dcolumn}
\usepackage{bm}
\usepackage{hyperref}
\usepackage{euscript}

 \newcommand{\eqre}[1]{Eq. \eqref{#1}}

\begin{document}


\title{Soft-gluon Resummation for High-$p_T$ Inclusive-Hadron Production at COMPASS}

\author{Daniel \surname{de Florian}}
\affiliation{
Departamento de F\'{i}sica, FCEyN, Universidad de Buenos Aires, (1428) Pabell\'{o}n 1, Ciudad Universitaria, Capital Federal, Argentina
}
\author{Melanie \surname{Pfeuffer}}
\author{Andreas \surname{Sch\"{a}fer}}
\affiliation{
 Institute for Theoretical Physics, University of Regensburg, D-93040 Regensburg, Germany
}%
\author{Werner \surname{Vogelsang}}
\affiliation{%
 Institute for Theoretical Physics, T\"{u}bingen University, Auf der Morgenstelle 14, D-72076 T\"{u}bingen, Germany
}%


\begin{abstract}
We study the cross section for the photoproduction reaction $\gamma N \rightarrow h X$ in fixed-target scattering at COMPASS, where the 
hadron $h$ is produced at large transverse momentum. We investigate the role played by higher-order QCD corrections to the cross section. In particular we address large logarithmic ``threshold'' corrections to the rapidity dependent partonic cross sections, which we resum to all orders at next-to-leading
accuracy. In our comparison to the experimental data we find that the threshold contributions are large and improve the agreement between data and theoretical predictions significantly. 
\end{abstract}

\maketitle


\section{Introduction}
\label{sec:intro}

Photoproduction processes in fixed-target lepton-nucleon scattering are important probes of nucleon structure. 
Cross sections for high-transverse-momentum ($p_T$) final states typically receive sizable or 
even dominant contributions from the photon-gluon fusion subprocess $\gamma g\to q\bar{q}$, offering access
to the nucleon's gluon distribution that is otherwise hard to obtain in lepton scattering.
Notably, the fixed target lepton scattering experiment COMPASS at CERN uses 
the process $\gamma N \rightarrow h X$ (where $h$ denotes a high-$p_T$ final-state hadron) in polarized
scattering in order to determine the nucleon's gluon helicity distribution $\Delta g$~\cite{Silva:2011zz}. 
Earlier measurements were made by the SLAC E155 experiment~\cite{Anthony:1999ac}.
Such measurements may provide information complementary to that obtained in polarized $pp$-scattering at 
RHIC~\cite{Aschenauer:2013woa}.

\begin{figure}[!t]
\vspace*{10mm}
\includegraphics[width=8cm]{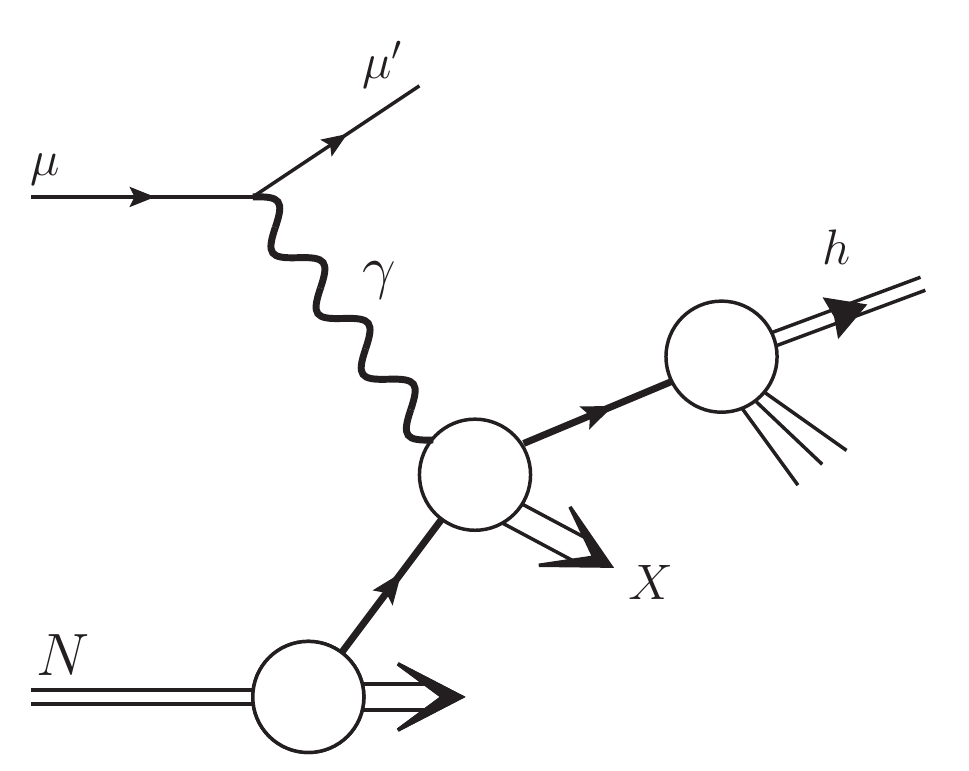}
\caption{Photoproduction in lepton-nucleon scattering. The virtual photon is required experimentally to have
low virtuality. \label{fig:photo}}
\end{figure}
At COMPASS, photoproduction $\gamma N \rightarrow h X$ is accessed in the lepton-nucleon scattering process 
$\mu N\to \mu' hX$ by selecting events with low virtuality of the exchanged photon (see Fig.~\ref{fig:photo}), 
typically $Q^2=-q^2\leq 0.1\,{\rm GeV}^2$. Such a selection is favored over deep-inelastic scattering at large $Q^2$
in terms of statistics since most of the 
events in lepton scattering are clustered at low $Q^2$. As long as the produced hadron's transverse momentum 
is large, the reaction can still be considered a hard-scattering reaction. From a theoretical point of view, however, the framework for 
the process $\gamma N \rightarrow h X$ in fixed-target scattering is relatively complex. While the radiation of the 
quasi-real photon from the incident lepton can be straightforwardly treated by the Weizs\"acker-Williams equivalent 
photon method, the interaction of the photon with the nucleon requires additional input as compared to 
usual deep-inelastic scattering. First, it is well known that a quasi-real photon does not always interact in an elementary
``direct'' way, but may also resolve into its own hadronic structure, described by parton distributions of the photon.
Although some of these resolved contributions also involve the nucleon's gluon distribution, they will overall
tend to dilute the sensitivity of photoproduction to the gluon density somewhat. 
For unpolarized photons, measurements at HERA and LEP have provided a fair amount of information on the 
photon's parton distributions (for review, see~\cite{Klasen:2002xb}), so that the resolved components in the cross section may be computed 
relatively reliably. In the polarized case, very little is known about the photon's parton content. Estimates 
based on next-to-leading order calculations have shown here~\cite{Jager2005,Jager2003c} (see also~\cite{Afanasev:1997ie})
that the process remains a good probe of $\Delta g$ even in the presence of resolved contributions. 

The other reason why theoretical calculations are quite involved is due to the fixed-target kinematics employed in experiment.
Typically transverse momenta are such that the variable $x_T=2p_T/\sqrt{s}$ (with $\sqrt{s}$ the center-of-mass energy)
is relatively large, say, $x_T\gtrsim 0.1$. It turns out that the partonic hard-scattering cross sections relevant for $\gamma N \rightarrow h X$ 
are then largely probed in the ``threshold''-regime, where the initial photon and parton have just enough energy to produce the 
high-transverse momentum parton that subsequently fragments into the hadron, and its recoiling counterpart. 
Relatively little phase space is then available for additional radiation of partons. In particular, gluon radiation is inhibited and 
mostly constrained to the emission of soft and/or collinear gluons. The cancellation of infrared singularities between real and 
virtual diagrams then leaves behind large double- and single-logarithmic corrections to the partonic cross sections. 
These logarithms appear for the first time in the next-to-leading order (NLO) 
expressions for the partonic cross sections, where (for the rapidity-integrated
cross section) they arise as terms of the form $\alpha_s\ln^2(1-x_T^2)$, with $\alpha_s$ the strong coupling constant.
At yet higher ($k$th) order of perturbation theory, the double-logarithms are of the form $\alpha_s^k\ln^{2k}(1-x_T^2)$. 
When the threshold regime dominates, it is essential to take into account the large logarithms to all orders in the strong coupling, 
a technique known as ``threshold resummation''. For single-hadron production in $pp\to hX$ in the fixed-target regime,
the resummation has been carried out in~\cite{Florian2005a}, and substantial effects were observed that lead to an enhancement of
the cross section. The same was found for the related process $pp\to h_1 h_2X$~\cite{Almeida:2009jt}. It is therefore to be
expected that threshold resummation effects are also relevant for the process $\gamma N \rightarrow h X$. 

In the present paper we will address this issue and investigate the resummation effects on the spin-averaged cross section for
$\gamma N \rightarrow h X$ in the COMPASS kinematic regime. Resummation affects the direct and the resolved contributions 
to the cross section differently, which makes photoproduction a particularly interesting process from the point of view of
resummation. Our results are directly relevant for comparison to recent  COMPASS data~\cite{Adolph2012,Hoppner:2012owa}. 
We note that we also extend the previous work~\cite{Florian2005a} by including rapidity dependence in resummation, 
following the techniques developed in~\cite{Almeida:2009jt}. In our phenomenological studies we find that the threshold logarithms 
play an important role for COMPASS kinematics, yielding essential higher-order corrections. Thus, resummation turns out to 
be a vital ingredient for comparisons between theory and experimental data. 

The paper is organized as follows. In Sec. \ref{sec:framework} we recall the basic framework for photoproduction of a hadron. 
In Section \ref{sec:resummed} we present details of threshold resummation and describe the technique 
that enables us to get a resummed expression for fixed rapidity of the observed hadron. 
%
Section \ref{sec:phenomen_results} is devoted to a phenomenological analysis,
focusing on the kinematics of the COMPASS experiment. We briefly conclude in Sec. \ref{sec:Conclusion}.

\section{\label{sec:framework}Technical framework}

We consider the unpolarized cross section for the semi-inclusive process
\begin{equation}\label{eq:process}
\ell N \rightarrow \ell' h^\pm  X,
\end{equation}
where a lepton beam scatters off a nucleon target $N$ producing
a hadron $h$ with transverse momentum $p_T$ and pseudorapidity $\eta$ in the final state. The basic
concept that links the experimentally measurable quantities to
theoretical predictions made from perturbative calculations is the
factorization theorem. It states that large momentum-transfer
reactions may be split into long-distance pieces, 
the universal parton distribution functions, and short-distance
contributions reflecting the hard interactions of the partons. Thus
we may write the unpolarized rapidity dependent differential cross section for the
process in \eqre{eq:process} as the following convolution \cite{Collins1985,Sterman1987}:
\begin{align}\nonumber
  \frac{p_T^3 d \sigma}{d p_T d\eta} =\sum_{a,b,c} \int_{x_{\ell}^{{\mathrm{min}}}}^1
d x_\ell\int_{x_n^{{\mathrm{min}}}}^1 
d x_n  \int_{x}^1 dz \frac{\hat x_T^4 z^2}{8v}\\  \times f_{a/\ell}(x_\ell,\mu_{fi})  f_{b/N}(x_n,
\mu_{fi})  D_{h/c}(z,\mu_{ff})
\frac{\hat s d\hat\sigma_{ab\rightarrow cX}}{dv dw}\,.\label{eq:factorTheorem}
\end{align}
The sum in \eqre{eq:factorTheorem} extends
over all possible partonic channels with $\hat\sigma_{ab\rightarrow cX}$ denoting the associated partonic
hard scattering cross section. In addition to the renormalization scale
$\mu_r$, the factorization of the hadronic cross section requires the
introduction of two further scales: the factorization scales $\mu_{fi}, \mu_{ff}$ for the initial and final states, respectively. All scales are
arbitrary but should be of the order of the hard scale. One usually chooses them to be equal, typically 
$\mu_r=\mu_{fi}=\mu_{ff}=p_T$. 
The parton distributions of the lepton and the
nucleon, $f_{a/\ell}(x_{\ell},\mu_{fi}),f_{b/N}(x_n,\mu_{fi})$, are evolved to the factorization scale
and depend on the respective momentum fractions 
$x_{\ell,n}$ carried by partons $a$ and $b$. $D_{h/c}(z,\mu_{ff})$ denotes the parton-to-hadron fragmentation function.
The lower bounds in the integrations over the various momentum fractions in \eqre{eq:factorTheorem} read: 
\begin{equation}
x_{\ell}^{{\mathrm{min}}}=\frac{x_Te^{\eta}}{2-{x_T}e^{-\eta}},~
x_n^{{\mathrm{min}}}=\frac{x_Te^{-\eta}}{2 -\frac{x_T}{x_\ell} e^\eta},~
x= \frac{x_T\cosh\hat\eta}{\sqrt{x_n x_\ell}}.
\end{equation}
Here $\hat \eta$ and $\hat x_T$ are the partonic counterparts to the pseudorapidity $\eta$ and the hadronic scaling variable $x_T=2p_T/\sqrt{s}$,
\begin{equation}
 \hat\eta=\eta + \frac 1 2 \ln {\frac{x_n}{x_\ell}}, ~ \hat x_T=\frac{x_T}{z \sqrt{x_\ell x_n}}\,.
\end{equation}
It is common convention to introduce two variables $v$ and $w$,
\begin{equation}
 v=1-\frac{\hat x_T}{2}e^{-\hat\eta}, ~ w= \frac{1}{v}\frac{\hat x_T}{2}e^{\hat\eta}\,,
\end{equation}
and to rewrite the partonic cross section in terms of this new set of variables.
Furthermore we introduce the
Mandelstam variables
\begin{align}\nonumber
 \hat s&= x_n x_\ell s, ~~\hat t = (p_a -p_c)^2=-\hat s \hat x_T e^{-\hat\eta}/2,\\
 \hat u &=(p_b-p_c)^2=-\hat s \hat x_T e^{\hat\eta}/2.
\end{align}
The invariant mass of the unobserved partonic final state is
\begin{align}
 s_4 = \hat s + \hat t + \hat u = \hat s v (1-w) = \hat s (1-\hat x_T \cosh \hat \eta).
\end{align}

The partonic hard-scattering functions $\hat\sigma_{ab\rightarrow cX}$ can be evaluated in QCD
perturbation theory. They may each be written as an expansion in the strong coupling constant $\alpha_s(\mu_r)$ of the
form
\begin{equation}
 \hat\sigma_{ab\rightarrow cX}(v,w)=\hat\sigma_{ab\rightarrow cX}^{(0)}(v,w)+{\alpha_s(\mu_r)}\hat\sigma_{ab\rightarrow cX}^{(1)}(v,w)+\EuScript{O}(\alpha_s^2).
\end{equation}

Whenever a photon takes part in a hard scattering process as initial particle,
one generally distinguishes two contributions, the so-called ``direct''
and ``resolved'' photon contributions,
\begin{equation}
 d\sigma=d\sigma_{\rm dir}+d\sigma_{\rm res}.
\end{equation}
Applying the Weizs\"acker-Williams equivalent photon method to the lepton-to-parton
distribution functions, $f_{a/\ell}$ in \eqre{eq:factorTheorem} may be written as a convolution
of a lepton-to-photon splitting function $P_{\gamma \ell}$ and a parton
distribution function $f_{a/\gamma}$ of a photon:
\begin{equation}\label{eq_partons_in_lepton}
 f_{a/\ell}(x_\ell,\mu_f)=\int_{x_\ell}^1 \frac{dy}{y} P_{\gamma
\ell}(y)f_{a/\gamma}(x_\gamma=\frac{x_\ell}{y},\mu_f).
\end{equation}
In the unpolarized case the splitting function is given by \cite{Frixione1993a,Florian1999a}
\begin{eqnarray}\label{eq_Weizswillspectrum}
\begin{aligned}
 P_{\gamma \ell}(y)=\frac{\alpha}{2\pi}\left[
\frac{1+(1-y^2)}{y}\ln\frac{Q^2_{\rm max}(1-y)}{m_\ell^2 y^2}\right. \\
+\left. 2 m_\ell^2 y \left( \frac{1}{Q^2_{\rm max}}-\frac{1-y}{m_\ell^2
y^2}\right)\right],
\end{aligned}\end{eqnarray}
and describes the collinear emission of a quasi-real photon with momentum
fraction $y$ off a lepton $\ell$ of mass $m_\ell$. The virtuality of the radiated photon
is restricted to be less than $Q_{\rm max}$, which is in turn constrained by the
experimental setup.

In the direct case, the photon participates as a whole and parton $a$ in
\eqre{eq:factorTheorem} is an elementary photon. Consequently, we here have simply 
\begin{equation}
 f_{\gamma/\gamma}(x_\gamma,\mu_f)=\delta(1-x_\gamma).
\end{equation}
There are two basic partonic subprocesses in lowest order (LO), in which a photon and a
parton from the initial nucleon give rise to the production of a hadron: 
photon-gluon-fusion $\gamma g\to q\bar{q}$ and Compton scattering $\gamma q\to q g$.
For each process, either of the final-state partons may hadronize into the observed hadron. 
As the processes are partly electromagnetic and partly due to strong interaction their cross 
sections are proportional to $\alpha\alpha_s(\mu_r)$, where $\alpha$ represents the electromagnetic fine structure
constant.

In addition to that the photon exhibits also a hadronic structure in the
framework of QCD. This is  described by the resolved process. 
Unlike
hadronic parton distributions, photonic densities may be decomposed into a
purely perturbatively calculable ``pointlike'' contribution and a nonperturbative
``hadron-like'' part. While the pointlike contribution dominates at large momentum fractions 
$x_\gamma$, the latter dominates in the low-to-mid $x_\gamma$ region and may be estimated
via the vector-meson-dominance model \cite{Schuler1996,Gluck1999}. 
At lowest order there are the following resolved subprocesses:
\begin{align}
& qq'\rightarrow qq',~~ q\bar q'\rightarrow q\bar q',~~q\bar q\rightarrow q'\bar q',~~qq\rightarrow qq,~~q\bar q\rightarrow q\bar q, \nonumber\\
& q\bar q\rightarrow gg,~~ gq\rightarrow qg, ~~ qg\rightarrow gq, ~~ gg\rightarrow gg, ~~ gg\rightarrow q\bar q.  
\end{align}
Each of these is a pure QCD-process and therefore has a cross section 
quadratic in $\alpha_s(\mu_r)$. However, as the photon parton distributions are 
formally of order $\alpha/\alpha_s(\mu_f)$, the perturbative expansion of the direct and resolved contributions 
starts at the same order. 

At LO where one has $2\rightarrow 2$ kinematics, $w\equiv 1$, and therefore,
\begin{equation}
 \frac{\hat s d \hat \sigma^{(0)}_{ab\rightarrow cX}(v,w)}{dv dw}=\frac{\hat s d \hat \sigma^{(0)}_{ab\rightarrow cd}(v)}{dv}\delta(1-w).
\end{equation}
The numerous partonic NLO cross sections $\hat\sigma^{(1)}_{ab\rightarrow cX}(v,w)$ have been computed 
in \cite{Aversa1989,Jager2003}. They can be cast into the form
\begin{align}
 \frac{\hat s d \hat \sigma^{(1)}_{ab\rightarrow cX}(v,w)}{dv dw}=&A(v)\delta(1-w) + B(v) \left(\frac{\ln(1-w)}{1-w} \right)_+ \nonumber\\ &+C(v)\left(\frac{1}{1-w} \right)_+ + F(v,w). \label{eq_NLO}
\end{align}
Here the ``+''-distributions are defined as follows:
\begin{equation}
 \int_0^1 f(w) [g(w)]_+ dw = \int_0^1 [f(w)-f(1)]g(w) dw. 
\end{equation}
The function $F(v,w)$ collects all remaining terms that do not contain any distributions. 
The terms in \eqre{eq_NLO} associated with ``+''-distributions yield large logarithmic first order corrections close to the threshold. These terms can be traced back to soft gluon emission and will also show up in all higher order corrections. For each new order of perturbations theory one is faced with two more powers of  leading logarithmic contributions. To be specific, in the $k$th order in perturbation theory $d\hat \sigma^{(k)}_{ab\rightarrow cX}(v,w)/dv dw$ contains logarithms of the form $\alpha_s^k[\ln^{2k-1}(1-w)/(1-w)]_+$, plus subleading terms with fewer logarithms.
Depending on kinematics, these logarithmic terms have to be resummed order-by-order.

\section{\label{sec:resummed}Resummed cross section}
In this section we will provide the resummed differential cross section as a function of transverse momentum $p_T$ and pseudorapidity $\eta$ of the produced hadron.

\subsection{Mellin moments and threshold region \label{MM}}
Threshold resummation of soft gluon emissions is performed in Mellin-$N$ moment space. Taking  Mellin moments transforms 
a convolution of a parton distribution function and the partonic cross section into a product of moments of the corresponding quantities. 
The  threshold region $w\rightarrow1$ corresponds to large Mellin moments. Under this transformation, the large soft-gluon corrections  
showing up as ``+''-distributions are translated into powers of logarithms $\ln N$.  This logarithmic behavior colludes with 
the $N$-dependence of the parton distribution functions and the fragmentation function, which in moment 
space typically fall off as $1/N^4$ or faster at large $N$. 

The single-inclusive cross section we are interested in here depends on two kinematic variables, $p_T$ and $\eta$. 
If the cross section is integrated over all rapidities, it becomes a function of $x_T^2$, and a single Mellin moment
in $x_T^2$ suffices to factorize it in terms of moments of parton distributions, fragmentation functions, and partonic
cross sections~\cite{Florian2005a}. After resummation the full Mellin expression 
is inverted, directly giving the desired hadronic cross section. 
If, on the other hand, one is interested in the rapidity dependence of the resummed
cross section, the integrations of the various functions in~\eqre{eq:factorTheorem} are no longer convolutions in a strict 
sense, and a different technique needs to be used. A convenient possibility~\cite{Almeida:2009jt} is to use Mellin moments for
only a part of the terms in~\eqre{eq:factorTheorem}. That is, one takes Mellin moments only of the product of fragmentation
functions and the resummed partonic cross sections, performs a Mellin inverse, and convolutes the result with the 
parton distributions in $x$-space. Inclusion of the fragmentation functions in the Mellin moment expression guarantees that
the integrand for the inverse Mellin transform falls off fast enough for the integral to show good numerical convergence. 
On the other hand, performing the convolution with the parton distributions in $x$-space provides full control over
rapidity, since the partonic and hadronic rapidities are related by a boost along the collision axis that involves only 
the momentum fractions of the initial-state partons. 

To be specific, starting from \eqre{eq:factorTheorem}, we consider only the last integral and take moments
in $x^2$ (where $x$ is the lower bound of the $z$-integral). The integral then factorizes into a
product of moments:
\begin{eqnarray}
&&\int_0^1 dx^2 \,(x^2)^{N-1} 
\int_{x}^1 dz \frac{\hat x_T^4 z^2}{8v} D_{h/c}(z,\mu_{ff})
\frac{\hat s d\hat\sigma_{ab\rightarrow cX}}{dv dw}\nonumber \\[2mm]
&&\equiv\,D_{h/c}^{2N+3}(\mu_{ff})
\tilde w^{2N}(\hat \eta),\label{Dwprod}
\end{eqnarray}
where  the Mellin moments $D_{h/c}^{N}(\mu_{ff})$ of the fragmentation function are defined as usual by
\begin{equation}
 D_{h/c}^N(\mu_{ff})=\int_0^1 dz z^{N-1} D_{h/c}(z,\mu_{ff}),
\end{equation}
and where the hard scattering function $\tilde w^N(\hat \eta)$ is given in Mellin-$N$ moment space by
\begin{equation}
 \tilde w^N(\hat \eta)=2\int_0^1 d\frac{s_4}{\hat s}\left(1-\frac{s_4}{\hat s}\right)^{N-1}\frac{\hat x_T^4 z^2}{8v}\frac{\hat s d\hat\sigma_{ab\rightarrow cX}}{dv dw}.
\end{equation}
We next take the Mellin inverse of the expression in~\eqre{Dwprod} which is then convoluted with the parton distributions: 
\begin{align}\nonumber
  \frac{p_T^3 d \sigma}{d p_T d\eta} =&\sum_{a,b,c}  \int_{0}^1
d x_\ell \int_{0}^1 d x_n  f_{a/\ell}(x_\ell,\mu_{fi}) f_{b/N}(x_n,\mu_{fi})\\  
&\times\int_{\cal{C}}\frac{d N}{2\pi i} (x^2)^{-N} D_{h/c}^{2N+3}(\mu_{ff})
\tilde w^{2N}(\hat \eta).\label{eq:strategy}
\end{align}
This is mathematically equivalent to using~\eqre{eq:factorTheorem} in the first place. However, 
once one uses a resummed hard-scattering function, it is much better from a computational point of
view to use the procedure in~(\ref{eq:strategy}) since the moments of the fragmentation functions
tame the large-$N$ behavior of the $\tilde w^{2N}$ so that the Mellin integral converges rapidly. 
In contrast, to carry out a convolution over $z$ as in \eqre{eq:factorTheorem} would become 
very difficult for a resummed hard-scattering function, since the latter contains ``+''-distributions with any power of
a logarithm. 


\subsection{Rapidity-dependent resummation to next-to-leading logarithm}

\begin{figure}[!t]
\includegraphics[width=9cm]{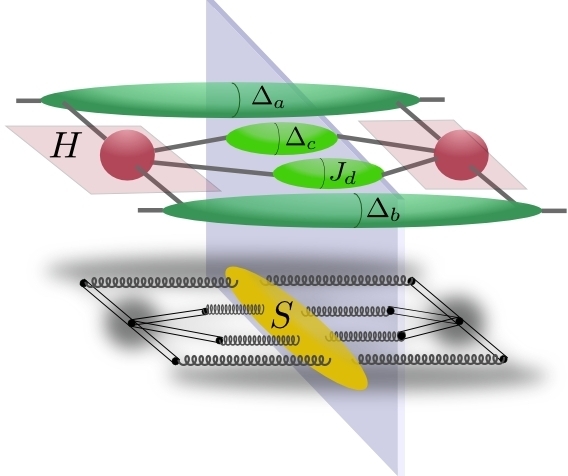}
\caption{Schematic illustration of the factorization of the cross section close to threshold. 
The plane symbolizes the cut, which separates the amplitude and its complex conjugte.
For each of the initial- and final-state partons there is a
function ($\Delta_{a,b,c}, J_d$) describing soft-gluon emissions off these partons. The hard-scattering part $H$ is depicted as amplitude
and its complex conjugate. It can be completely separated from the soft function $S$, shown in the lower part of the figure. The double lines represent eikonal lines. \label{fig:factorization}}
\end{figure}

It turns out that multigluon QCD amplitudes factorize to logarithmic accuracy. Furthermore, in Mellin space, 
also the phase space including the constraint of energy conservation factorizes.
The resummed cross section in moment space factorizes into functions for each single participating parton, 
a function describing the hard scattering, and a soft function. This factorization is illustrated in Fig.\,\ref{fig:factorization}. 
The resummed cross section is given by \cite{Kidonakis1997,Kidonakis1998a,Kidonakis1998,Bonciani:2003nt}:
\begin{align}
 \tilde w^{{\mathrm{resum}}, N}_{ab\rightarrow cd}(\hat\eta)=&\Delta_a^{N_a}(\hat s,\mu_{fi})\Delta_b^{N_b}(\hat s,\mu_{fi})\Delta_c^{N}
 (\hat s,\mu_{ff})\nonumber\\
&\times J_d^{N}(\hat s)   {\mathrm{Tr}}\left\{ H \EuScript{S}_{N}^\dagger S \EuScript{S}_{N}\right\}_{ab\rightarrow cd},\label{eq:w_resum}
\end{align}
where $N_a=(-\hat u/\hat s)N$ and $N_b=(-\hat t/\hat s)N$. 
The resummed exponents for the initial-state partons ${a,b=q,\bar{q},g}$ 
in \eqre{eq:w_resum} read, in the $\overline{\text{MS}}$ scheme:
\begin{align}\nonumber
 \ln \Delta_i^{N}(M_i^2, \mu_{fi}) =& -\int_0^1 dz \frac{z^{N-1}-1}{1-z}\nonumber\\
 &\times\left\{\int_{(1-z)^2}^1 \frac{dt}{t} A_i\left[ \alpha_s\left(t M_i^2\right)\right]\right.\nonumber \\
&+\left.\bar B_i\left(\nu_i,\frac{M_i^2}{\hat s},\alpha_s\left((1-z)^2M_i^2\right) \right)\right\} \nonumber\\
&- 2\int_{\mu_r}^{M_i}\frac{d\mu'}{\mu'}\gamma_i\left(\alpha_s(\mu'^2)\right)\nonumber\\
&+ 2\int_{\mu_{fi}}^{M_i}\frac{d\mu'}{\mu'}\gamma_{ii}\left(N,\alpha_s(\mu'^2)\right).\label{eq:Delta_initial}
\end{align}
Here $M_i$ is a scale of order $\sqrt{\hat s}$. It was shown that the exponent is in fact independent of $M_i$ 
at next-to-leading logarithmic (NLL) accuracy \cite{Sterman2001}. Furthermore, we have 
\begin{align}
& A_i(\alpha_s)=C_i[{\alpha_s}/{\pi}+ ({K} /2)  \left({\alpha_s}/{\pi}\right)^2], \nonumber \\
& \bar B_i(\nu_i, M_i^2/\hat s, \alpha_s)=C_i(\alpha_s/\pi)[1-\ln(2\nu_i)+\ln(M_i^2/\hat s)],\label{AB}
\end{align}
where $K=C_A[{67}/{18}-{\pi^2}/{6}]- 5/9 N_f$, $C_i=C_q=C_F=4/3$ for an incoming quark, and $C_i=C_g=C_A=3$ for a gluon. $N_f$ denotes the number of flavors. The $\nu_i$ are defined as
\begin{equation}
 \nu_i\equiv \frac{(\beta_i\cdot n)^2}{|n^2|},
\end{equation}
with the parton velocity $\beta_i^\mu=p^\mu_i\sqrt{2/\hat s}$ and an axial gauge vector $n$. The $\nu_i$ were introduced to make 
the factorization of the cross section manifest \cite{Laenen1998}. The gauge-dependence they express 
will cancel in the final resummed cross section.  The last two terms in \eqre{eq:Delta_initial} match the exponent to the chosen renormalization and factorization scale, respectively. The $\gamma_i$ are the anomalous dimensions of the quark and gluon fields, and the $\gamma_{ii}$ 
correspond to the logarithmic and constant terms of the moments of the diagonal Altarelli-Parisi splitting functions. To one loop order, one has
\begin{align}
 &\gamma_q(\alpha_s)=\frac{3}{4} C_F \frac{\alpha_s}{\pi},~~ \gamma_{qq}(N,\alpha_s)=-\left(\ln N -\frac 3 4\right) C_F \frac{\alpha_s}{\pi},\nonumber\\
&\gamma_g(\alpha_s)=b_0 \alpha_s,~~~~~~ \gamma_{gg}(N,\alpha_s)=-\left(C_A\ln N -\pi b_0\right)\frac{\alpha_s}{\pi},
\end{align}
where $b_0=4\pi\beta_0=(11 C_A-4 T_R N_f)/(12\pi)$, with $T_R=1/2$ and the one-loop coefficient of the $\beta$-function, $\beta_0$. 
We note that the large-$N$ behavior of the diagonal splitting functions and anomalous dimensions links the various terms in the exponent
in~\eqre{eq:Delta_initial} to each other,
\begin{align}
 \gamma_{ii}(N,\alpha_s)= -\ln\bar N A_i(\alpha_s) + \gamma_i(\alpha_s),
\end{align}
 where $\bar N=N e^{\gamma_E}$ with the Euler constant $\gamma_E$. 
 
For the direct processes, parton $a$ is a photon and we have simply $\Delta_\gamma^{N_a}(\hat s, \mu_{fi})=1$.
For the fragmenting parton $c$ one has the same exponent as for the incoming partons $\Delta_i^N(M_i^2)$ in \eqre{eq:Delta_initial}, 
but  with the final state factorization scale $\mu_{ff}$ in place of the initial-state one.

The exponential function $J_d^N$ in \eqre{eq:w_resum} contains collinear emission, both soft and hard, by the unobserved 
final-state jet that recoils against the observed parton. It is independent of factorization scale and is given by
\begin{align}\nonumber
 \ln J_d^N(\hat s)=&\int_0^1 dz \frac{z^{N-1}-1}{1-z}\left\{\int_{(1-z)^2}^{(1-z)}\frac{dt}{t}A_d[\alpha_s(t\hat s)]\right.\\
 &-\gamma_d[\alpha_s((1-z)\hat s)] -\bar B_d[\nu_d,1,\alpha_s((1-z)^2 \hat s)]{\bigg\}} \nonumber\\
 &+2\int_{\mu_r}^{\sqrt{\hat s}}\frac{d\mu'}{\mu'}\gamma_d(\alpha_s(\mu'^2)),
\end{align}
where $A_d$, $\gamma_d$ and $\bar B_d$ are defined as in \eqre{AB}.

Finally, coherent soft gluon radiation among the jets is treated by the last term in \eqre{eq:w_resum}. The functions $H_{ab\rightarrow cd}$, $\EuScript{S}_{N,{ab\rightarrow cd}}$ and $S_{ab\rightarrow cd}$ are matrices in a space of color exchange operators \cite{Kidonakis1997,Kidonakis1998a,Kidonakis1998}, and the trace is taken in this color space. The $H_{ab\rightarrow cd}$ are the hard-scattering functions. They are perturbative series in $\alpha_s$,
\begin{align}
 H_{ab\rightarrow cd} (\hat\eta, \alpha_s)=H_{ab\rightarrow cd}^{(0)} (\hat\eta) + \frac{\alpha_s}{\pi}H_{ab\rightarrow cd}^{(1)} (\hat\eta)+\EuScript{O}(\alpha_s^2).
\end{align}
The LO contributions to the hard-scattering functions in the resolved-photon case 
are known with their full color dependence \cite{Kidonakis1997,Kidonakis1998a,Kidonakis1998,Kidonakis2001},
and the NLO terms have been obtained in~\cite{Kelley:2010fn,Catani:2013vaa}. 
The $S_{ab\rightarrow cd}$ are soft functions and may be expanded as
\begin{align}
 S_{ab\rightarrow cd}(\hat \eta,\alpha_s)=S_{ab\rightarrow cd}^{(0)}+\frac{\alpha_s}{\pi}S_{ab\rightarrow cd}^{(1)}(\hat \eta,\alpha_s, \frac{\sqrt{\hat s}}{N})+\EuScript{O}(\alpha_s^2).
\end{align}
Here, the Mellin-$N$ moment enters only in the argument of the running coupling \cite{Kidonakis1997}. Therefore, the $N$-dependence of the soft functions will show up at next-to-next-to-leading logarithmic order 
for the first time. The LO terms $S_{ab\rightarrow cd}^{(0)}$ for the resolved contribution may be taken from \cite{Kidonakis1997,Kidonakis1998a,Kidonakis1998}, while the $S_{ab\rightarrow cd}^{(1)}$ are not yet available in
closed form. Contributions by soft gluons emitted at wide angles 
are resummed by 
the exponentials $\EuScript{S}_{ab\rightarrow cd}(\hat \eta,\alpha_s)$, which are evolved via the soft anomalous dimension matrices $\Gamma_{ab\rightarrow cd}$:
\begin{align}\label{softad}
 \EuScript{S}_{N,ab\rightarrow cd}(\hat \eta,\alpha_s)=\EuScript{P}
\exp\left[\int_{\mu_r}^{\sqrt{\hat s}/N}\frac{d\mu'}{\mu'}\Gamma_{ab\rightarrow cd}(\hat\eta,\alpha_s(\mu'))\right],
\end{align}
with $\EuScript{P}$ denoting path ordering, and the soft anomalous dimensions expanded as follows:
\begin{align}\label{softad1}
 \Gamma_{ab\rightarrow cd}(\hat\eta,\alpha_s)=\frac{\alpha_s}{\pi}\Gamma_{ab\rightarrow cd}^{(1)}(\hat\eta)+\EuScript{O}(\alpha_s^2).
\end{align}
The first-order terms may be found in \cite{Kidonakis1997,Kidonakis1998a,Kidonakis1998} and have the structure
\begin{align}
 \left(\Gamma_{ab\rightarrow cd}^{(1)}(\hat\eta)\right)_{mn}&=\left(\tilde\Gamma_{ab\rightarrow cd}^{(1)}(\hat\eta)\right)_{mn}\nonumber\\
 &+\delta_{mn}\sum_{k=a,b,c,d} \frac{C_{k}}{2}[-\ln(2\nu_k)+1-\pi i],
\end{align}
where one may see the gauge-dependent diagonal elements explicitly. As mentioned before, gauge-dependence cancels 
in the above expressions for the resummed cross section to next-to-leading-logarithmic accuracy. 

Let us take a look at the first-order expansion of the trace part in \eqre{eq:w_resum} (see~\cite{Almeida:2009jt}):
\begin{align}
& {\mathrm{Tr}} 
\left\{ H \EuScript{S}_{N}^\dagger S \EuScript{S}_{N} \right\}_{ab\rightarrow cd} = {\mathrm{Tr}}\left\{ H^{(0)} S^{(0)} \right\}_{ab\rightarrow cd} \nonumber\\
& +\frac{\alpha_s}{\pi} {\mathrm{Tr}}\{ -[H^{(0)}(\Gamma^{(1)})^\dagger S^{(0)}+H^{(0)} S^{(0)}\Gamma^{(1)}]\ln  N   \nonumber\\ 
& + H^{(1)} S^{(0)}+H^{(0)} S^{(1)}\}_{ab\rightarrow cd} +\EuScript{O}(\alpha_s^2).
\end{align}
The trace of the product of the matrices $H$ and $S$ at lowest order reproduces the Born cross sections. As discussed in
\cite{Almeida:2009jt,Catani:2013vaa}, in order to 
obtain ${\mathrm{Tr}} \left\{ H \EuScript{S}_{N}^\dagger S \EuScript{S}_{N} \right\}$ fully to NLL accuracy one would need to 
implement the contributions from $H^{(1)}$ and $S^{(1)}$, which is
beyond the scope of this work. Following the approach of \cite{Florian2005a,Almeida:2009jt}, we use the approximation
\begin{align}
 {\mathrm{Tr}} \left\{ H \EuScript{S}_{N}^\dagger S \EuScript{S}_{N} \right\}_{ab\rightarrow cd} \approx & \left(1+\frac{\alpha_s}{\pi}C_{ab\rightarrow cd}^{(1)} \right) \nonumber\\
 & \times {\mathrm{Tr}} \left\{ H^{(0)} \EuScript{S}_{N}^\dagger S^{(0)} \EuScript{S}_{N} \right\}_{ab\rightarrow cd},\label{eq:approximation_Ccoeff}
\end{align}
where the so-called ``$C$-coefficients'' are defined as
\begin{align}
 C_{ab\rightarrow cd}^{(1)}(\hat\eta)\equiv\frac{{\mathrm{Tr}}\{H^{(1)}S^{(0)}+H^{(0)}S^{(1)}\}_{ab\rightarrow cd}}{\mathrm{Tr}\{H^{(0)}S^{(0)}\}}.
\end{align}
This approximation becomes exact for color-singlet cases, and therefore in particular for the direct subprocesses which have only one color 
structure at Born level. The $C$-coefficients are constructed in such a way that the first order expansion of the resummed 
cross section reproduces all terms $\propto \delta(1-w)$ in the NLO result.

\subsection{Rapidity-dependent NLL exponents}
The expression for the resummed partonic cross section in \eqre{eq:w_resum} is formally ill-defined for any value of $N$, as its exponents involve integrations of the running coupling over the Landau pole. However, it was shown that the divergencies showing up in \eqre{eq:w_resum} are subleading in $N$ \cite{Catani1996a}.
In return, a NLL expansion of the resummed formula is finite up to $N$ reaching the first Landau pole at 
$N_L=\exp(1/(2\alpha_s b_0))$. We will return to this point later. 
We now rewrite the resummed exponents for soft gluon radiation off the 
incoming and outcoming partons in \eqre{eq:w_resum} as expansions to NLL accuracy using the perturbative expansions
given in~(\ref{AB}):
\begin{align}
A_i(\alpha_s)&=\frac{\alpha_s}{\pi}A_i^{(1)}+\left(\frac{\alpha_s}{\pi}\right)^2 A_i^{(2)}+\EuScript{O}(\alpha_s^3),\\
B_i(\alpha_s)&=\frac{\alpha_s}{\pi} B_i^{(1)}+\EuScript{O}(\alpha_s^2), \\ 
\bar B_i(\alpha_s)&=\frac{\alpha_s}{\pi} \bar B_i^{(1)}+\EuScript{O}(\alpha_s^2), 
 \end{align}
 where $B_i^{(1)}=-2\gamma_i^{(1)}$ with $\gamma_i(\alpha_s)=\gamma_i^{(1)}\alpha_s/\pi+\EuScript{O}(\alpha_s^2)$.
The resulting exponents do not depend on the specific subprocess, but only on the type of parton and thus may be seen in this sense as 'universal' functions. The leading 
terms in the exponent are leading logarithms (LL) of the form $\alpha_s^k\ln^{k+1} N$, while subleading terms are down  at least 
by one power of $\ln N$. We adopt the formalism of \cite{Catani1998a} and organize the logarithms in the exponentials in a way such that all leading logarithmic terms are collected in functions $h_i^{(1)}$ and $f_i^{(1)}$ for the observed and the unobserved partons, respectively. 
These functions are rapidity independent and hence are identical to the analogous functions in the rapidity-integrated exponents. 
Rapidity dependent terms first appear at NLL accuracy, where they yield additional terms when compared to 
the well-known rapidity integrated exponents of \cite{Florian2005a}. 

We further expand the resummed exponents for the  observed partons $i=a,b,c$ and unobserved partons $d$ to NLL accuracy:
\begin{align}\nonumber
 \Delta_{i}^{2N_{i}}&(\hat\eta,\hat s,\mu_{fi})= \ln N h_{i}^{(1)}(\lambda)+h_{i}^{(2)}(\lambda,\frac{\hat{s}}{\mu_r^2},\frac{\hat{s}}{\mu_{fi}^2})\\
 &-\frac{A_{i}^{(1)}}{\pi b_0}\ln\left(\frac{2 N_{i}}{N}\right)\ln(1-2\lambda)-\frac{\bar B_{i}^{(1)}}{2\pi b_0}\ln(1-2\lambda),
\label{eq:Delta_i}
%
\\
 J_{d}^{2N}(\hat s)&= \ln N f_{d}^{(1)}(\lambda)+f_{d}^{(2)}(\lambda,\frac{\hat{s}}{\mu_r^2})-\frac{\bar 
 B_{d}^{(1)}}{2\pi b_0}\ln(1-2\lambda)\nonumber\\ 
&- \frac{A_{d}^{(1)}}{\pi b_0}\ln(2)\left(\ln(1-\lambda)-\ln(1-2\lambda)\right),
\label{eq:J_i}
\end{align}
where $\lambda=\alpha_s b_0\ln N$ and, as before, $N_a=(-\hat u/\hat s)N$ and $N_b=(-\hat t/\hat s)N$. 
For the observed final-state parton we simply have $N_c=N$. 
Note that due to the NLL expansion of terms like $\ln(1-\alpha_s b_0 \ln(N_i))\approx \ln(1-2\lambda)-\frac{2\alpha_s b_0}{1-2\lambda}\ln\left(N_i/N \right)$ explicit dependence on $\hat\eta$ appears in \eqre{eq:Delta_i}. 
The functions $h_{i}^{(k)},f_{i}^{(k)}$ are known from resummation for the rapidity-integrated cross sections and
are given by
\begin{align}
 h_i^{(1)}&(\lambda)=  \frac{A_i^{(1)}}{2\pi b_0 \lambda}\left[ 2\lambda + (1-2\lambda)\ln(1-2\lambda) \right],
\end{align}
\begin{align}
 h_i^{(2)}(\lambda,&\frac{Q^2}{\mu_r^2},\frac{Q^2}{\mu_a^2})=  -\frac{A_i^{(2)}}{2\pi^2 b_0^2}\left[2\lambda+\ln(1-2\lambda)\right]\nonumber\\
&- \frac{A_i^{(1)}\gamma_E}{\pi b_0}\ln(1-2\lambda) - \frac{A_i^{(1)}}{\pi b_0}\lambda \ln \frac{Q^2}{\mu_a^2}\nonumber\\\nonumber
 & + \frac{A_i^{(1)}b_1}{2\pi b_0^3}\left[2\lambda + \ln(1-2\lambda)+\frac 1 2 \ln^2(1-2\lambda)\right] \\
 & + \frac{A_i^{(1)}}{2\pi b_0}[2\lambda + \ln(1-2\lambda)]\ln \frac{Q^2}{\mu_r^2},
\end{align}
 and for the unobserved final-state parton
\begin{align}
 f_i^{(1)}(\lambda)=&- \frac{A_i^{(1)}}{2\pi b_0 \lambda}\left[ (1-2\lambda)\ln(1-2\lambda) \right.\nonumber\\
 &\left.-2(1-\lambda)\ln(1-\lambda)\right],
 \end{align}
\begin{align}
 \nonumber
 f_i^{(2)}&(\lambda,\frac{Q^2}{\mu_r^2})=  -\frac{A_i^{(1)}b_1}{2\pi b_0^3}\bigg[\ln(1-2\lambda) -2\ln(1-\lambda) \\ 
 & \left. + \frac 1 2 \ln^2(1-2\lambda)-\ln^2(1-\lambda)\right]+ \frac{B_i^{(1)}}{2\pi b_0}\ln(1-\lambda)\nonumber\\\nonumber
 &  - \frac{A_i^{(1)}\gamma_E}{\pi b_0}\left[\ln(1-\lambda)-\ln(1-2\lambda)\right]\\\nonumber
 & - \frac{A_i^{(2)}}{2\pi^2 b_0^2}\left[2\ln(1-\lambda)-\ln(1-2\lambda) \right]  \\
 & +\frac{A_i^{(1)}}{2\pi b_0}[2\ln(1-\lambda) - \ln(1-2\lambda)]\ln \frac{Q^2}{\mu_r^2}.
\end{align}
As before, $b_0=(11C_A-4 T_R N_f)/12 \pi$, and
\begin{align}
 b_1=\frac{1}{24\pi^2}\left( 17 C_A^2-5C_A N_f-3C_F N_f \right),
\end{align}
correspond to the first two coefficients of the QCD $\beta$-function.

The path-ordered matrix exponentiation of the soft anomalous dimension contribution in Eqs.~(\ref{softad}),(\ref{softad1}) 
proceeds as described in~\cite{Almeida:2009jt}. We use a numerical approach, iterating the exponentiation
to a very high order. Finally, when all terms in the exponent are combined, the LL terms $\alpha_s^k \ln^{k+1} N$ 
and the NLL terms $\alpha_s^k \ln^k N$ in the exponent of \eqre{eq:w_resum} reproduce the three towers of logarithms 
$\alpha_s^k\ln^{2k} N$, $\alpha_s^k\ln^{2k-1} N$, and $\alpha_s^k\ln^{2k-2} N$ in the cross sections, up to the
approximation concerning the $C$-coefficients discussed earlier. The $C$-coefficients for the direct part are given in
the next subsection. As those for the resolved part are rather lengthy, we do not present them here; they can be obtained upon request.  

\subsection{The direct contribution\label{sec:direct}}

Our discussion so far directly applies to the resolved-photon contributions. In the direct case, the resummation
framework simplifies thanks to the fact that the LO processes have only three colored particles and hence only
one specific color configuration. Nevertheless, a few remarks about the resummation for the direct part are in order, 
since this case has not been discussed in the previous literature in any detail. 

For the direct processes the hard-scattering functions $H_{\gamma b\rightarrow cd}$, 
the soft functions $S_{\gamma b\rightarrow cd}$, and the anomalous 
dimensions are scalars in color space. This allows us to simplify \eqre{eq:w_resum}:
\begin{align}
 & \tilde w^{{\mathrm{resum},N}}_{\gamma b\rightarrow cd}(\hat\eta)=\left(1+\frac{\alpha_s}{\pi}C^{(1)}_{\gamma b\rightarrow cd}\right)\Delta_b^{N_b}(\hat s,\mu_{fi})\Delta_c^{N}(\hat s,\mu_{ff})\nonumber\\
&\times  J_d^{N}(\hat s) \hat\sigma^{(0)}_{\gamma b\rightarrow cd}(N,\hat\eta) \nonumber\\
&\times   
\exp\left[\int_{\mu_r}^{\sqrt{\hat s}/N} \frac{d \mu'}{\mu'} 2 Re\Gamma_{\gamma b\rightarrow cd}(\hat \eta,\alpha_s(\mu'))\right],
\label{eq:w_resum_direct}
\end{align}
where we have defined the Mellin-$N$ moment of the Born cross sections as
\begin{equation}
\hat \sigma^{(0)}_{\gamma b\rightarrow cd}(N,\hat\eta) \equiv 2\int_0^1 d\frac{s_4}{\hat s}\left(1-\frac{s_4}{\hat s}\right)^{N-1}\frac{\hat x_T^4 z^2}{8v}\frac{\hat s d\hat\sigma^{(0)}_{ab\rightarrow cX}}{dv dw}.
\end{equation}
The partonic Born cross sections for the three direct processes are given by
\begin{align}
 &\frac{\hat s d\hat\sigma_{\gamma q\rightarrow qg}^{(0)}(v)}{\pi\alpha\alpha_s e_q^2dv}=\frac{\hat s d\hat\sigma_{\gamma q\rightarrow gq}^{(0)}(1-v)}{\pi\alpha\alpha_s e_q^2dv}= 2 C_F \frac{1+(1-v)^2}{1-v},
\\
&\frac{\hat s d\hat\sigma_{\gamma g\rightarrow q\bar q}^{(0)}(v)}{\pi\alpha\alpha_s e_q^2dv}= \frac{v^2+(1-v)^2}{v(1-v)}.
\end{align}
The soft anomalous dimensions for the direct processes may be derived from those for 
the prompt-photon production processes $qg\rightarrow \gamma q$, and 
$q\bar q\rightarrow \gamma g$ \cite{Sterman2001,Kidonakis2000, Laenen1998,Catani1998a,Catani1999}. 
The rapidity-dependent anomalous dimensions then read to first order:
\begin{align}
 \Gamma_{\gamma q\rightarrow q g}^{(1)}(\hat \eta)&=\frac{C_F}{2}  \left[2\ln\left(\frac{-\hat u}{\hat s}\right)-\ln(4\nu_{q_a}\nu_{q_c})+2 \right]\nonumber\\ 
&+ \frac{C_A}{2}\left[\ln\left(\frac{\hat t}{\hat u}\right)-\ln(2\nu_g)+1-\pi i\right],
\\
%
\Gamma_{\gamma q \rightarrow gq}^{(1)}(\hat \eta)&=\left.\Gamma_{\gamma q\rightarrow q g}^{(1)}(\hat \eta)\right|_{\hat t\leftrightsquigarrow \hat u},
\\
 \Gamma_{\gamma g\rightarrow q \bar q}^{(1)}(\hat \eta)&=\frac{C_F}{2} \Big[-\ln(4\nu_q\nu_{\bar q})+2-2\pi i\Big] \nonumber\\ 
&+\frac{C_A}{2}\left[\ln\left(\frac{\hat t\hat u}{\hat s^2}\right)+1-\ln(2\nu_g)+\pi i\right].
\end{align}
With the first order terms of the anomalous dimensions at hand, the integral in \eqre{eq:w_resum_direct} can be written explicitly 
as an expansion to NLL accuracy:
\begin{align}
 \int_{\mu_r}^{\sqrt{\hat s}/2N} \frac{d \mu'}{\mu'} 2 Re\Gamma_{\gamma b\rightarrow cd}(\hat \eta,\alpha_s(\mu'))=\frac{\Gamma_{\gamma b\rightarrow cd}^{(1)}(\hat\eta)}{\pi b_0}\ln(1-2\lambda).
\end{align}

We recall that with only one color configuration present at Born level, the approximation in \eqre{eq:approximation_Ccoeff} 
becomes exact, and the $C$-coefficients for the direct processes may be derived by comparing the exact NLO calculation \cite{Jager2003c}
to the first-order expansion of~\eqre{eq:w_resum_direct}. Moreover, it can be checked that all double- and single-logarithmic 
terms $\alpha_s\ln^2 N$, $\alpha_s\ln N$ (including the rapidity-dependence of the latter) 
are correctly reproduced by the resummation formula. For quark production via Compton scattering, one finds:
\begin{align}
 &C_{\gamma q \rightarrow q g}(\hat\eta) = \frac  {C_F} {2(1-v)}  \Bigg\{ 
(C_A-2C_F)
\nonumber\\
&\left.\left[ (3-2v)\left(\ln^2\frac{1-v}{v}+\pi^2
\right)\right.\right.\nonumber\\
&+\left.\left.(1-2v)\ln^2(v)+2(1-v)\ln\frac{1-v}{v^2}
\right] +6C_F\ln (1-v)
\right\} \nonumber\\
&+ \frac 1 4 S_{\gamma g} \left\{
C_F \left[
       6\ln^2\frac{1-v}{v} + \frac 1 2
(\rho_{q\gamma}^{(F)})^2  - 6\ln[v(1-v)]
\right.\right.
\nonumber\\
&+\rho_{q\gamma}^{(F)} \left( 3-2\ln[v(1-v)] \right)+8\ln v\ln(1-v)
 \nonumber \\ 
& \left.\left. 
  +
\rho_{q\gamma}^{(F)}\ln\frac{\mu_{fi}^2}{\hat s} + [-3+4(\gamma_E+\ln 2)] \ln
\frac{\mu_{ff}^2}{\hat s}+  \frac{16}{3} \pi^2
- \frac{19}{2}
\right]\right. \nonumber\\
&\left.
-C_A \left[
3\ln^2 \frac{1-v}{v}+
\frac{1}{8} (\rho_{q\gamma}^{(A)})^2   + \rho_{q\gamma}^{(A)} \ln\frac{1-v}{v} +
2\pi^2 -\frac 4 3
\right]
\right.\nonumber\\
&\left.
+4 \pi b_0 \left(\frac 5 3 + \gamma_E + \ln 2 +\ln \frac{\mu_{r}^2}{\hat
s}\right)
\right\},
\end{align}
where $S_{\gamma g}=2 C_F(1+(1-v)^2)/({1-v})$ and
\begin{align}
 \rho_{q\gamma}^{(F)}&=-3+4(\gamma_E + \ln[2(1-v)]), \\
 \rho_{q\gamma}^{(A)} &= 4 (\gamma_E + \ln 2). 
\end{align}
The $C$-coefficients are subject to LO-kinematics and therefore we have
\begin{align}
  v&=1+\frac{\hat t }{\hat s}=\frac{e^{\hat\eta}}{2\cosh\hat\eta}.
\end{align}
The $C$-coefficient for the production of a gluon, which then fragments into the observed hadron, reads:
\begin{align}
 &C_{\gamma q \rightarrow gq}(\hat\eta) = \frac  {C_F} {2v}  \left\{ 
(C_A-2C_F)\left[
(1+2v)\left(\pi^2 +
\ln^2\frac{v}{1-v}\right)
\right.\right.
\nonumber\\
& \left.\left.
-(1-2v)\ln^2(1-v)+2v\ln\frac{v}{(1-v)^2}
\right] +6C_F\ln v
\right\} \nonumber\\
&+ \frac 1 4 {\tilde S_{\gamma g}} \left\{
C_F \left[
       4\ln^2\frac{1-v}{v} +2(\ln(1-v)-3)\ln(1-v)
\right.\right. \nonumber\\
&+\left.
\rho_{q\gamma}^{(F)}\ln\frac{\mu_{fi}^2}{\hat s (1-v)} + \frac 1 8
(\rho_{q\gamma}^{(F)})^2 + \frac 3 2 \rho_{q\gamma}^{(F)} + \frac{11}{3} \pi^2
- \frac{29}{8}
\right] \nonumber\\
&\left.
-C_A \left[
\ln^2 \frac{1-v}{v} - \rho_{q\gamma}^{(A)} \ln \frac{\mu_{ff}^2}{\hat s} -
\frac{1}{4} (\rho_{q\gamma}^{(A)})^2\right.\right.\nonumber\\
&\left.\left. - \rho_{q\gamma}^{(A)} \ln\frac{1-v}{v} +
\frac{\pi^2}{3}
\right] \right. \nonumber\\
&\left.
-4 \pi b_0 \ln \frac{\mu_{ff}^2}{\mu_r^2}
\right\},
\end{align}
where $\tilde S_{\gamma g}=2 C_F (1+v^2)/{v}$. 
Finally, for the photon-gluon fusion process, one finds
\begin{align}
 &C_{\gamma g\rightarrow q \bar q}(\hat\eta) = -\frac  {1} {8v(1-v)}  \left\{ 
(C_A-2C_F)\left[ (1+2v)\ln^2 v
\right.\right.
\nonumber\\
& \left.
+(3-2v)\ln^2 (1-v) +  2\ln[v(1-v)]
\right]
\nonumber\\
& +6C_F (1-2v)\ln \frac{1-v}{v}
+ \frac 1 4 S_{\gamma g} \left\{ -4 \pi b_0 \ln \frac{\mu_{fi}^2}{\mu_r^2}
\right.
\nonumber\\
&
+C_F \left[\frac 1 8
(\rho_{g\gamma}^{(F)})^2 +\ln[v(1-v)]\left( 1+\ln[v(1-v)] \right)
\right.
\nonumber\\
& \left.
- 2\ln(1-v)\ln v 
+  \frac 3 2
\rho_{g\gamma}^{(F)} +
\rho_{g\gamma}^{(F)}\ln\frac{\mu_{ff}^2}{\hat s} + \frac{5}{3} \pi^2
- \frac{29}{8}
\right]
 \nonumber\\
&
+C_A \left[
\frac 1 2 \ln^2 [v (1-v)] + \ln[v(1-v)]\left(1- \rho_{g\gamma}^{(A)} \right)
\right.\nonumber\\
& \left.\left.
-\ln v\ln(1-v) + \rho_{g\gamma}^{(A)}
\ln  \frac{\mu_{fi}^2}{\hat s} +
\frac{1}{4} (\rho_{g\gamma}^{(A)})^2   +
\frac 2 3 \pi^2
\right]
\right\},
\end{align}
where now $S_{\gamma g}=\left(v^2+(1-v)^2\right)/(v(1-v))$ and
\begin{align}
 \rho_{g\gamma}^{(F)}&=-3+4(\gamma_E + \ln2), \\
 \rho_{g\gamma}^{(A)} &= 4 (\gamma_E + \ln [2(1-v)]).
\end{align}
\subsection{Inverse Mellin Transform and Matching Procedure}
\begin{figure}[!t]
\includegraphics[width=8cm]{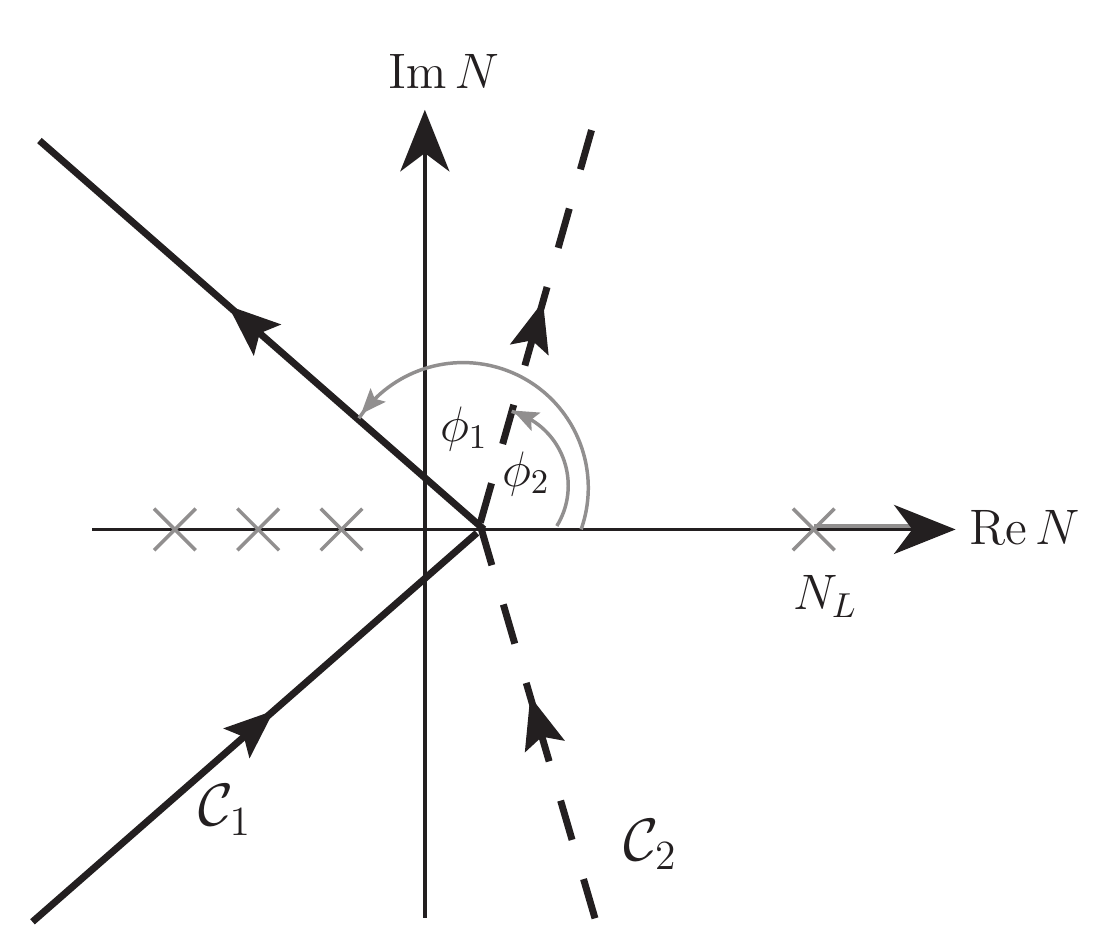}
\caption{The two contours in Mellin-$N$ space for inverting the product of moments of the resummed partonic cross sections and the fragmentation functions. The crosses symbolize the poles of the fragmentation functions and LO cross sections on the real axis. $N_L$ is the position of the leftmost Landau pole. \label{fig:contour}}
\end{figure}
Resummation takes place in Mellin-$N$ moment space, and one therefore needs an inverse Mellin transform to translate the 
result back into the physical space. As described in Sec.~\ref{MM} (see \eqre{eq:strategy}),
our approach has been to place the Mellin-$N$ transformation in between the convolutions over the parton distribution functions and the 
fragmentation and hard scattering functions. Therefore, the inverse Mellin transform that we need 
is given by
\begin{align}
 \sigma_D(x,\hat\eta)\equiv\int_{\cal{C}}\frac{d N}{2\pi i} (x^2)^{-N} D_{h/c}^{2N+3}(\mu_{ff})
\tilde w^{{\mathrm{resum}},2N}(\hat \eta).\label{eq:Mellininv}
\end{align}
The NLL expanded forms, Eqs.\,\eqref{eq:Delta_i},\,\eqref{eq:J_i}, have singularities for $\lambda=1/2$ and $\lambda=1$, known as Landau poles 
and corresponding to moments $N_L=\exp (1/(2\alpha_s b_0))$ and $N_L=\exp (1/(\alpha_s b_0))$, respectively, that are
located on the positive real axis in moment space. 
Therefore a prescription has to be found for dealing with these singularities. We follow the \emph{minimal prescription} \cite{Catani1996a},
according to which the contour for the inverse transformation 
runs between the first Landau pole $N_L$ and the rightmost of all other poles of the integrand. This choice ensures that the 
perturbative expansion is an asymptotic series that has no factorial divergence \cite{Catani1996a}.
Because of the branch cuts starting at the Landau poles to the right of the contour, the inverted $\sigma_D(x,\hat\eta)$ has 
support at $x>1$ \cite{Catani1996a,Almeida:2009jt}. Although the contribution from this unphysical region
decreases exponentially with $x$, we find that it is not negligible for the kinematics of interest for phenomenology,
even after subsequent convolution with the parton distributions. This possibly points to significant non-perturbative
effects for the cross section and kinematic regime we consider here. 

For our numerical computations, we choose the inverse Mellin contours ${\cal{C}}_1$ (for $x<1$) and ${\cal{C}}_2$ (for $x>1$)
illustrated in Fig.\,\ref{fig:contour} in the complex-$N$ plane. Bending the contours at non-zero angles with respect to the 
imaginary axis improves the numerical convergence of the integrals. The contour ${\cal{C}}_2$ is still chosen to be rather steep,
in order to avoid strong oscillations resulting from the branch cuts. 

When using resummation to provide theoretical predictions of cross sections, one wants to make use of the best fixed-order
theoretical calculation available, which in this case is NLO. 
Therefore, we ``match'' our resummed cross section to the NLO one. This is achieved by expanding the 
partonic cross sections to the first non-trivial order in $\alpha_s$ ($\EuScript{O}(\alpha_s^2)$ for the direct
case, $\EuScript{O}(\alpha_s^3)$ for the resolved one), subtracting the expanded result from the resummed one, 
and adding the full NLO cross section:
\begin{align}
 \nonumber
&  \frac{p_T^3 d \sigma^{{\mathrm{matched}}}}{d p_T d\eta} =\sum_{a,b,c}  \int_{0}^1
d x_\ell \int_{0}^1 d x_n   \nonumber \\
&\times f_{a/\ell}(x_\ell,\mu_{fi})f_{b/N}(x_n,\mu_{fi})\nonumber\\[2mm]
&\times\int_{\cal{C}}\frac{d N}{2\pi i} (x^2)^{-N} D_{h/c}^{2N+3}(\mu_{ff})\nonumber \\[2mm]
&\times
\left[\tilde w^{{\mathrm{resum}}, 2N}_{ab\rightarrow cd}(\hat \eta) -\left.\tilde w^{{\mathrm{resum}}, 2N}_{ab\rightarrow cd}(\hat \eta)\right|_{
{\mathrm{first-order}}} \right]\nonumber\\[2mm]
&+\frac{p_T^3 d \sigma^{{\mathrm{NLO}}}}{d p_T d\eta}\label{eq:res_match}.
\end{align}
This procedure allows to take into account the NLO calculation in full. The soft-gluon contributions beyond NLO are resummed to NLL. 

\section{\label{sec:phenomen_results}Phenomenological Results}
\begin{figure}[t]
\includegraphics[width=9cm]{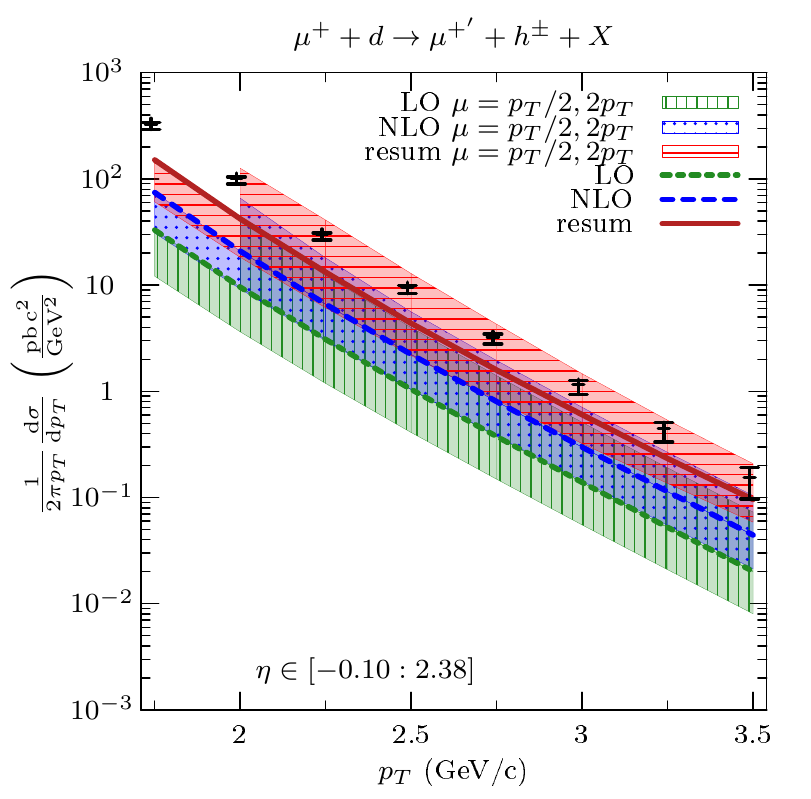}
\caption{Comparison of the LO, NLO and resummed (``resum'') calculations to COMPASS data. The error bars shown for 
the experimental data are the quadratic sums of statistical and systematic uncertainties. In addition, there is a 10\% normalization uncertainty due
to the luminosity determination. For scale 
$\mu=p_T/2$ we only show results for $p_T\geq 2\,$GeV. \label{fig:cross_LO_NLO_res}}
\end{figure}
Starting from \eqre{eq:res_match} we now compare the resummed cross section to experimental hadron production data measured at the COMPASS experiment at CERN \cite{Adolph2012}. In this fixed-target experiment muons at a beam energy of $E_\mu=160\,$GeV were scattered off a 
deuteron target, corresponding to a lepton-nucleon center-of-mass energy of $\sqrt{s}=17.4\,$GeV.  Due to a detector area cut the fraction $y$ of the lepton momentum carried by the photon is restricted to the range $0.2\leq y\leq 0.8$. For the COMPASS photoproduction studies the maximally allowed 
virtuality $Q^2_{{\mathrm{max}}}$ of the photons was $Q^2_{{\mathrm{max}}}=
0.1\,{\rm GeV}^2$. The measured hadrons $h^\pm$ were subject to the following kinematic cuts: the fraction $z_{{\mathrm{cut}}}$ 
of the virtual photon energy carried by the detected hadron had to be within the range $0.2\leq z_{{\mathrm{cut}}} \leq 0.8$. 
In addition, the scattering angle $\theta$ of the observed hadron was constrained by $10\leq \theta \leq 120\,$mrad, corresponding to 
$2.38\geq\eta\geq -0.1$ in pseudo-rapidity in the lepton-nucleon center-of-mass system. 

In our calculations we use the CTEQ6M5 set of parton distribution functions  for the nucleon \cite{Tung2007} and the ``Gl\"{u}ck-Reya-Schienbein''
(GRS) parton distribution functions of the photon \cite{Gluck1999}. For the fragmentation functions we use the ``de\,Florian-Sassot-Stratmann'' (DSS) 
set \cite{Florian2007}. 
All scales in \eqre{eq:res_match} are set equal, $\mu=\mu_r=\mu_{fi}=\mu_{ff}=p_T$. In order to investigate the 
scale dependence of our results, we will also show the results for  $\mu=p_T/2$ and $\mu=2 p_T$.

\begin{figure}[t]
\includegraphics[width=9cm]{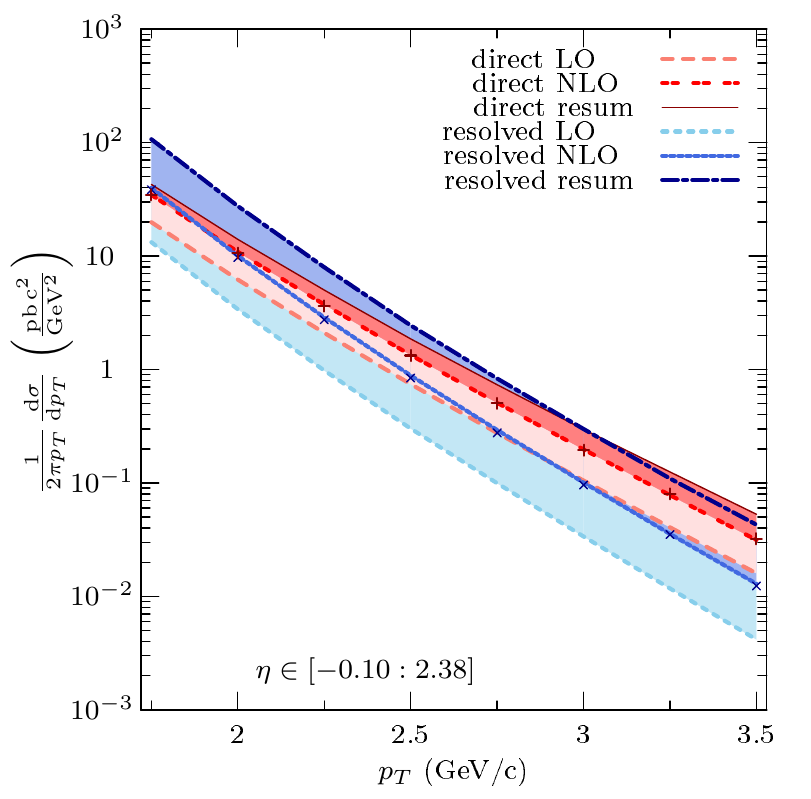}
\caption{The direct contribution versus the resolved one in the 
$\overline{\text{MS}}$ scheme for the photon's parton distributions. 
The crosses denote the first-order expansions of the direct and resolved resummed cross sections.\label{fig:direct-resolved}}
\end{figure}
\begin{figure}[!t]
\includegraphics[width=9cm]{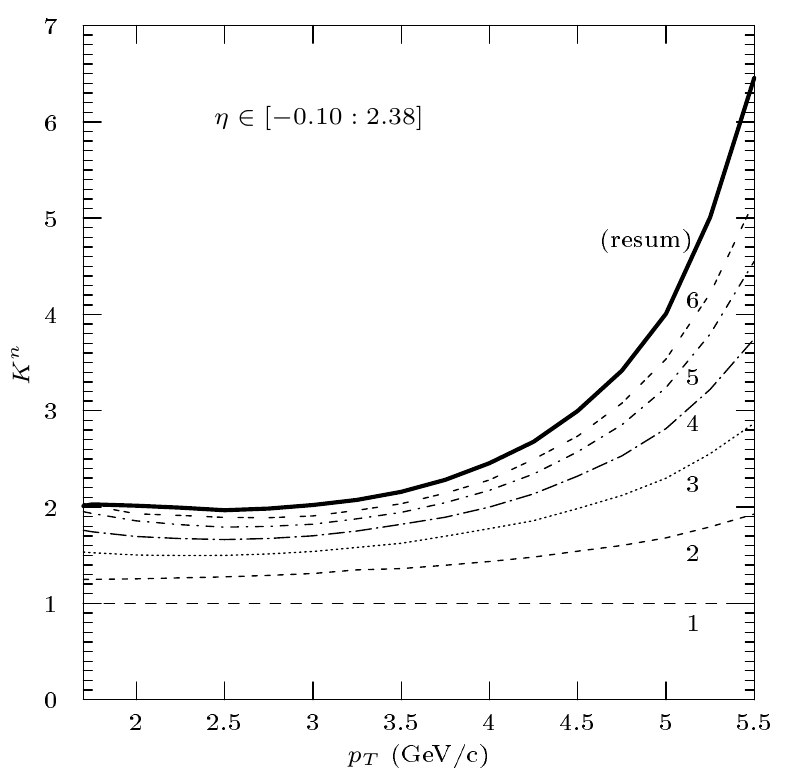}
\caption{Soft-gluon ``$K$-factors'' relative to NLO, as defined in \eqre{eq:def_Kfactor}, for 
COMPASS kinematics. The numbers labeling the curves refer to the superscript $n$ in $K^{(n)}$ in Eq.~(\ref{eq:def_Kfactor}). 
\label{fig:Kfactors}}
\end{figure}

In Fig.\,\ref{fig:cross_LO_NLO_res} we present our results for the matched resummed cross section for photoproduction in $\mu d\to h^\pm X$
for COMPASS kinematics and compare it to the experimental data \cite{Adolph2012}. Note that the data are available down to low transverse momentum $p_T=1.2\,$GeV, while we start our theoretical cross sections at $p_T=1.75\,$GeV to make sure that application of perturbative methods
is sensible. 
For all our calculations we have applied the cuts on the momentum fraction $y$ in the Weizs\"{a}cker-Williams photon and on
the photon's maximal virtuality given above. Moreover, thanks to our rapidity-dependent resummed approach, we are 
able to take into account the proper pseudo-rapidity cuts  $2.38\geq\eta\geq -0.1$ as well as $0.2\leq z_{{\mathrm{cut}}} \leq 0.8$ directly. 
Fig.\,\ref{fig:cross_LO_NLO_res} also shows  the LO and the NLO cross section. One observes that the LO one is far below the data. 
The  NLO corrections are huge, which indicates the importance of going beyond NLO and taking into account the threshold logarithms to all orders. 
The matched resummed cross section gives again a sizeable correction to the NLO result, enhancing the latter by a factor of about two. 
One observes that the resummed results agree with the data within the (admittedly, large) systematic error. Note that unfortunately 
for the kinematics discussed here the scale uncertainty of the resummed result is not really smaller than that of the LO or the NLO one.  

Even if neither the direct contribution $d\sigma_{{\mathrm{dir}}}$ nor the resolved one $d \sigma_{{\mathrm{res}}}$ are individually measurable quantities as both of them depend on the scheme chosen for the factorization of singular collinear parton emissions, it is 
instructive to consider both parts separately. The direct processes will generally dominate at high $p_T$. On the other hand,
in contrast to the direct processes, the resolved 
ones have an additional intermediate particle generated by the photon. As this carries only a fraction $x_\gamma$ of the photon momentum,  
less phase space is available for producing a high-momentum hadron. Therefore the resolved processes are on average 
closer to the partonic threshold, and thus we expect the threshold logarithms to have more impact than for the direct contribution. 
In addition, the resolved processes involve four colored partons, making them more likely to radiate soft gluons. Fig.\,\ref{fig:direct-resolved}
compares the direct and resolved contributions and the resummation effects on them. 
At lowest order the direct contribution exceeds the resolved one over the whole $p_T$-range considered. This changes already 
at NLO: Because of 
the large size of the NLO corrections in the resolved case, the resolved NLO cross section exceeds the direct NLO one 
at $p_T \leq 2\,$GeV. This trend continues for the resummed cross sections. 

In order to see whether the large effects from soft-gluon resummation correctly give the dominant part of the cross section, we 
perform a consistency check. For each subprocess the resummed cross section (not the matched one) is expanded to NLO and 
compared to the corresponding full fixed-order NLO result. We find that these expansions reproduce the NLO results very well 
for all processes, except for $gg\rightarrow q \bar q$ which, however, only makes a small contribution to the full cross section.
We have not been able to identify the reason for the discrepancy in this particular case, except that we found that it is
due to terms not related to ``+''-distributions. Figure \,\ref{fig:direct-resolved} also shows these comparisons, 
again separately for the direct and resolved contributions, where for each of the two we have combined all relevant subprocesses.
As can be observed, the agreement of the 
expansion and the NLO result is excellent. This implies that the terms that are formally suppressed by an 
inverse power of the Mellin moment $N$ near threshold indeed are insignificant. Thus one may safely assume that this will 
also be the case for higher-order corrections, so that the resummed cross section yields a good approximation to the all-order 
perturbative cross section. 

\begin{figure}[t]
\includegraphics[width=9cm]{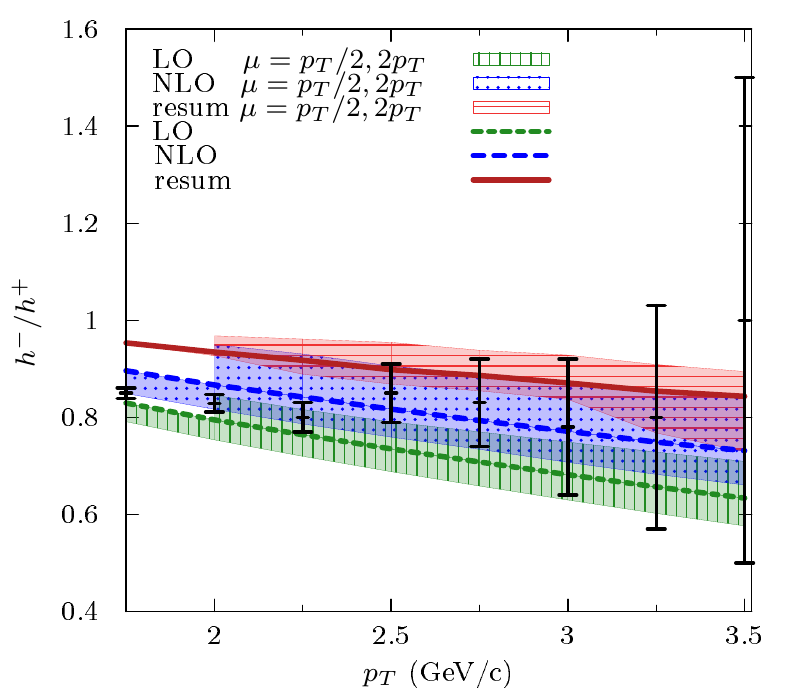}
\caption{Ratio of production cross sections for $h^-$ over $h^+$: 
$d\sigma_{\mu^+ d\rightarrow \mu^{+'} + h^-}/d\sigma_{\mu^+ 
d\rightarrow \mu^{+'} + h^+}$. The error bars of the experimental data are statistical only. \label{fig:chargeratio}}
\end{figure}

\begin{figure}[!t]
\includegraphics[width=9cm]{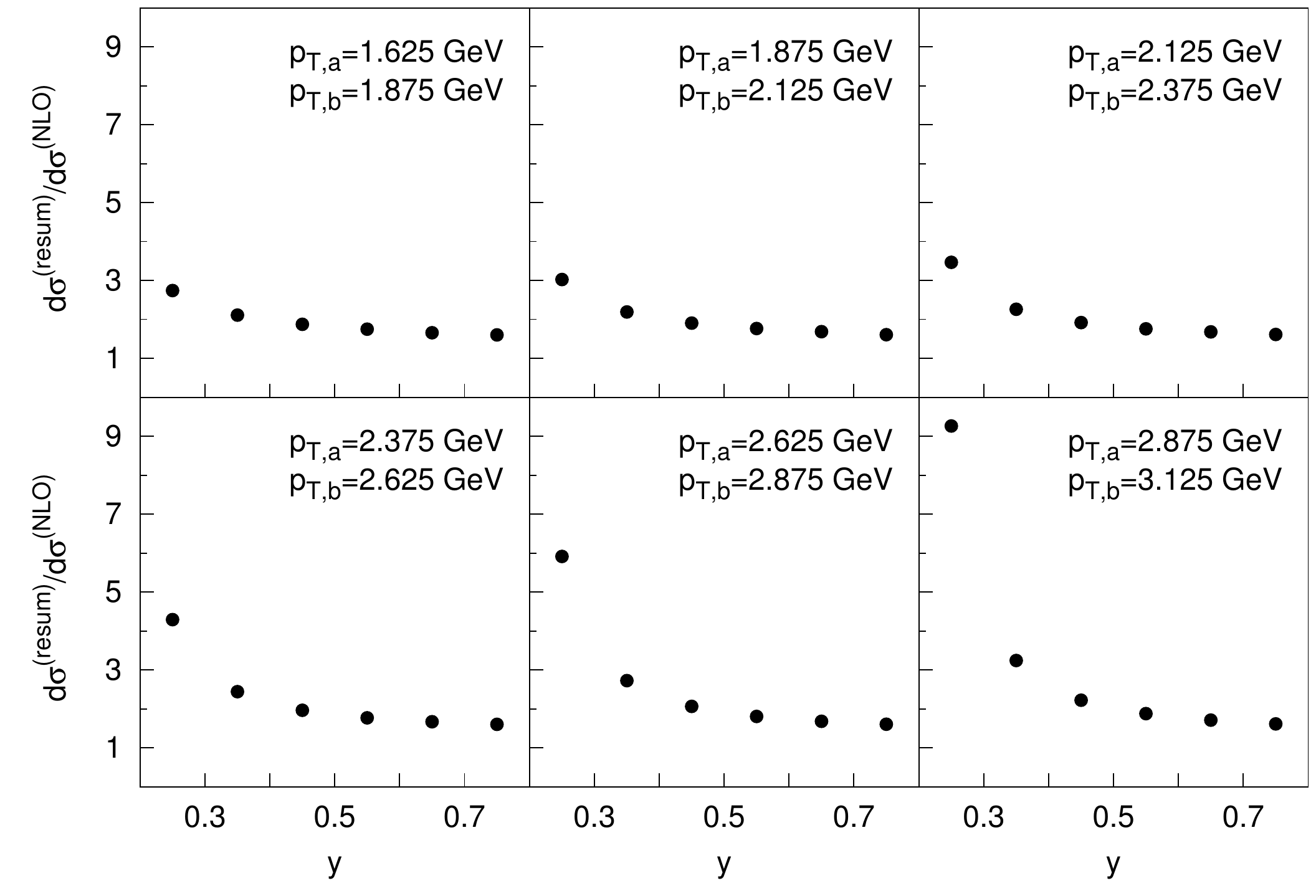}
\caption{Ratios of the $y$-dependent resummed and NLO cross sections, integrated over various
$p_T$ bins. The cross sections are averaged over the ranges $[y-0.05:y+0.05]$. \label{fig:y-dep}}
\end{figure}

We now investigate how the large enhancement of the NLL resummed cross section that we observed in Fig.\,\ref{fig:cross_LO_NLO_res} 
builds up order by order. We therefore expand the matched resummed formula beyond NLO and define the ``soft-gluon $K$-factors''
\begin{align}
 K^{(n)}\equiv \frac{\left.d\sigma^{{\mathrm{matched}}}/d p_T\right|_{\EuScript{O}(\alpha_s^{1+n})}}{d\sigma^{{\mathrm{NLO}}}/dp_T}
 \label{eq:def_Kfactor}.
\end{align}
In addition, $K^{{\mathrm{resum}}}\equiv K^{(\infty)}$ is defined as the ratio of the matched resummed cross section to the NLO one. Because of
the matching procedure given by \eqre{eq:res_match}, the first-order expansion of the matched resummed cross section is 
identical to the full fixed order NLO result, and we have $K^{(1)}=1$. Figure \,\ref{fig:Kfactors} shows $K^{{\mathrm{resum}}}$ along 
with the six lowest soft-gluon $K$-factors. One can see that they are almost flat for $p_T\leq 3.5\,$GeV but exhibit 
a dramatic enhancement for higher transverse momenta. Figure \,\ref{fig:Kfactors} also indicates that the series 
$K^{(1)}, K^{(2)}, K^{(3)} \ldots$ converges towards $K^{{\mathrm{resum}}}$, which may be regarded 
as further evidence for the importance of resummation. 

Next, we study the ratio of the production cross section for negatively charged hadrons over the one 
for positively charged hadrons. This ratio is also accessible at COMPASS. Figure \,\ref{fig:chargeratio} 
shows our calculation compared to the data. As expected, the production of positively charged mesons is preferred. 
This effect mostly stems from the QCD-Compton process $\gamma q\to  q g$ in the direct channel which
couples to up quarks four times as strongly as to down quarks. 
This tendency is most distinct for LO and softens when going to NLO and to the NLL resummed cross section, 
since resolved higher-order contributions are gaining importance. Figure \,\ref{fig:chargeratio} shows that the 
resummed cross section somewhat overpredicts the charge ratio measured in experiment. We note, however, that
we have obtained the scale uncertainty bands in the figure by simply dividing the $h^-$ and $h^+$ 
cross sections for a given scale. The true scale uncertainty on the ratio will likely be larger as one
could, in principle, choose different scales in the computation of the two cross sections.

Finally, in Fig.\,\ref{fig:y-dep} we investigate the dependence of the cross section on the photon energy 
fraction $y$ in the Weizs\"{a}cker-Williams spectrum. 
We consider the double differential cross section $d^2\sigma/(dp_T dy)$ integrated over $p_T$ bins
and averaged over the ranges $[y-0.05:y+0.05]$,
\begin{align}
\frac{1}{0.1}
\int_{y-0.05}^{y+0.05}dy'
\int_{p_T,a}^{p_T,b}\frac{d^2\sigma}{d p_T dy'} d p_T.
\end{align}
At fixed transverse momentum the phase space available for the production of additional partons is smaller, 
the smaller the photon energy fraction $y$. Therefore, for decreasing $y$ one gets closer to partonic threshold, and
one expects an increase of the cross section due to the impact of soft gluon emissions. 
This behavior is more pronounced at higher $p_T$. As $y$ is directly accessible in experiment, the 
$y$-dependence of the cross section may give information about whether hadron production at this kinematics is 
well-described by perturbative methods. 

\section{\label{sec:Conclusion}Conclusions}
We have studied the effects of next-to-leading logarithmic threshold logarithms on the direct- and resolved-photon cross sections for the process 
$\mu^+ d\rightarrow \mu^{+'} + h + X$ at high transverse momentum of the hadron $h$. As a new technical ingredient to resummation,
we were able to fully include the rapidity dependence of the cross section in the resummed calculation and to account for all 
relevant experimental cuts. This was achieved by treating only the partonic cross sections and the fragmentation functions in Mellin-$N$ 
moment space, but keeping the convolutions with the parton distribution functions in $x$-space. 

For COMPASS kinematics, we have found large higher-order soft-gluon QCD corrections. These are due to the fact
that one is overall rather close to the threshold region, as shown by the relatively large value of the hadron's
transverse momentum over the available center-of-mass energy,  typically $2p_T/\sqrt{s}\gtrsim 0.1$. 
The threshold logarithms addressed by resummation strongly dominate the higher-order corrections. We have verified this 
by comparing the first-order expansion of our resummed cross section with the full NLO one, finding excellent agreement of the two.
We have observed a significant enhancement of the resummed cross section over the next-to-leading order one, 
showing that the NLO calculations 
are likely not fully sufficient. Resummation also significantly improves the agreement between the data and theoretical predictions. 
It will be interesting to extend our calculations to the case of helicity asymmetries for this process, which are used at COMPASS 
to access the nucleon's spin-dependent gluon distribution.

\begin{acknowledgments}
We are grateful to C. H\"{o}ppner, B. Ketzer, C. Mar\-chand, A. Morreale, and many other members of COMPASS for constructive collaboration, and G. Sterman for useful discussions. 
This work was supported by the ``Bundesministerium f\"ur Bildung und Forschung'' (BMBF) (grants no. 06RY7195, 05P12WRFTE,
and 05P12VTCTG). M.P. was supported by a grant of the ``Studienstiftung des deutschen Volkes''. D.deF. was supported by UBACYT, CONICET, ANPCyT and the Research Executive Agency (REA) of the European Union under the Grant Agreement number PITN-GA-2010- 264564 (LHCPhenoNet). W.V. acknowledges support by Marie Curie 
Reintegration Grant IRG 256574 ResuQCD.

\end{acknowledgments}



\nocite{*}

\bibliography{references_hadronproduction.bib}

\begin{thebibliography}{34}
\expandafter\ifx\csname natexlab\endcsname\relax\def\natexlab#1{#1}\fi
\expandafter\ifx\csname bibnamefont\endcsname\relax
  \def\bibnamefont#1{#1}\fi
\expandafter\ifx\csname bibfnamefont\endcsname\relax
  \def\bibfnamefont#1{#1}\fi
\expandafter\ifx\csname citenamefont\endcsname\relax
  \def\citenamefont#1{#1}\fi
\expandafter\ifx\csname url\endcsname\relax
  \def\url#1{\texttt{#1}}\fi
\expandafter\ifx\csname urlprefix\endcsname\relax\def\urlprefix{URL }\fi
\providecommand{\bibinfo}[2]{#2}
\providecommand{\eprint}[2][]{\url{#2}}

\bibitem[{\citenamefont{Silva}(2011)}]{Silva:2011zz}
\bibinfo{author}{\bibfnamefont{L.}~\bibnamefont{Silva}}
  (\bibinfo{collaboration}{Compass Collaboration}), \bibinfo{journal}{PoS}
  \textbf{\bibinfo{volume}{EPS-HEP2011}}, \bibinfo{pages}{301}
  (\bibinfo{year}{2011}).

\bibitem[{\citenamefont{Anthony et~al.}(1999)}]{Anthony:1999ac}
\bibinfo{author}{\bibfnamefont{P.}~\bibnamefont{Anthony}} \bibnamefont{et~al.}
  (\bibinfo{collaboration}{E155 Collaboration}), \bibinfo{journal}{Phys.Lett.}
  \textbf{\bibinfo{volume}{B458}}, \bibinfo{pages}{536} (\bibinfo{year}{1999}),
  \eprint{hep-ph/9902412}.

\bibitem[{\citenamefont{Aschenauer et~al.}(2013)\citenamefont{Aschenauer,
  Bazilevsky, Boyle, Eyser, Fatemi et~al.}}]{Aschenauer:2013woa}
\bibinfo{author}{\bibfnamefont{E.}~\bibnamefont{Aschenauer}},
  \bibinfo{author}{\bibfnamefont{A.}~\bibnamefont{Bazilevsky}},
  \bibinfo{author}{\bibfnamefont{K.}~\bibnamefont{Boyle}},
  \bibinfo{author}{\bibfnamefont{K.}~\bibnamefont{Eyser}},
  \bibinfo{author}{\bibfnamefont{R.}~\bibnamefont{Fatemi}},
  \bibnamefont{et~al.} (\bibinfo{year}{2013}), \eprint{1304.0079}.

\bibitem[{\citenamefont{Klasen}(2002)}]{Klasen:2002xb}
\bibinfo{author}{\bibfnamefont{M.}~\bibnamefont{Klasen}},
  \bibinfo{journal}{Rev.Mod.Phys.} \textbf{\bibinfo{volume}{74}},
  \bibinfo{pages}{1221} (\bibinfo{year}{2002}), \eprint{hep-ph/0206169}.

\bibitem[{\citenamefont{J\"{a}ger et~al.}(2005)\citenamefont{J\"{a}ger,
  Stratmann, and Vogelsang}}]{Jager2005}
\bibinfo{author}{\bibfnamefont{B.}~\bibnamefont{J\"{a}ger}},
  \bibinfo{author}{\bibfnamefont{M.}~\bibnamefont{Stratmann}},
  \bibnamefont{and}
  \bibinfo{author}{\bibfnamefont{W.}~\bibnamefont{Vogelsang}},
  \bibinfo{journal}{Eur.Phys.J.} \textbf{\bibinfo{volume}{C44}},
  \bibinfo{pages}{533} (\bibinfo{year}{2005}), \eprint{hep-ph/0505157}.

\bibitem[{\citenamefont{J\"{a}ger
  et~al.}(2003{\natexlab{a}})\citenamefont{J\"{a}ger, Stratmann, and
  Vogelsang}}]{Jager2003c}
\bibinfo{author}{\bibfnamefont{B.}~\bibnamefont{J\"{a}ger}},
  \bibinfo{author}{\bibfnamefont{M.}~\bibnamefont{Stratmann}},
  \bibnamefont{and}
  \bibinfo{author}{\bibfnamefont{W.}~\bibnamefont{Vogelsang}},
  \bibinfo{journal}{Phys.Rev.} \textbf{\bibinfo{volume}{D68}},
  \bibinfo{pages}{114018} (\bibinfo{year}{2003}{\natexlab{a}}),
  \eprint{hep-ph/0309051}.

\bibitem[{\citenamefont{Afanasev et~al.}(1998)\citenamefont{Afanasev, Carlson,
  and Wahlquist}}]{Afanasev:1997ie}
\bibinfo{author}{\bibfnamefont{A.}~\bibnamefont{Afanasev}},
  \bibinfo{author}{\bibfnamefont{C.~E.} \bibnamefont{Carlson}},
  \bibnamefont{and}
  \bibinfo{author}{\bibfnamefont{C.}~\bibnamefont{Wahlquist}},
  \bibinfo{journal}{Phys.Rev.} \textbf{\bibinfo{volume}{D58}},
  \bibinfo{pages}{054007} (\bibinfo{year}{1998}), \eprint{hep-ph/9706522}.

\bibitem[{\citenamefont{de~Florian and Vogelsang}(2005)}]{Florian2005a}
\bibinfo{author}{\bibfnamefont{D.}~\bibnamefont{de~Florian}} \bibnamefont{and}
  \bibinfo{author}{\bibfnamefont{W.}~\bibnamefont{Vogelsang}},
  \bibinfo{journal}{Phys.Rev.} \textbf{\bibinfo{volume}{D71}},
  \bibinfo{pages}{114004} (\bibinfo{year}{2005}), \eprint{hep-ph/0501258}.

\bibitem[{\citenamefont{Almeida et~al.}(2009)\citenamefont{Almeida, Sterman,
  and Vogelsang}}]{Almeida:2009jt}
\bibinfo{author}{\bibfnamefont{L.~G.} \bibnamefont{Almeida}},
  \bibinfo{author}{\bibfnamefont{G.~F.} \bibnamefont{Sterman}},
  \bibnamefont{and}
  \bibinfo{author}{\bibfnamefont{W.}~\bibnamefont{Vogelsang}},
  \bibinfo{journal}{Phys.Rev.} \textbf{\bibinfo{volume}{D80}},
  \bibinfo{pages}{074016} (\bibinfo{year}{2009}), \eprint{0907.1234}.

\bibitem[{\citenamefont{Adolph et~al.}(2012)}]{Adolph2012}
\bibinfo{author}{\bibfnamefont{C.}~\bibnamefont{Adolph}} \bibnamefont{et~al.}
  (\bibinfo{collaboration}{COMPASS Collaboration}) (\bibinfo{year}{2012}),
  \eprint{1207.2022}.

\bibitem[{\citenamefont{H\"{o}ppner}(2012)}]{Hoppner:2012owa}
\bibinfo{author}{\bibfnamefont{C.}~\bibnamefont{H\"{o}ppner}}
  (\bibinfo{collaboration}{COMPASS Collaboration}) (\bibinfo{year}{2012}),
  \eprint{Technical U. Munich thesis, CERN-THESIS-2012-005}.

\bibitem[{\citenamefont{Collins et~al.}(1985)\citenamefont{Collins, Soper, and
  Sterman}}]{Collins1985}
\bibinfo{author}{\bibfnamefont{J.~C.} \bibnamefont{Collins}},
  \bibinfo{author}{\bibfnamefont{D.~E.} \bibnamefont{Soper}}, \bibnamefont{and}
  \bibinfo{author}{\bibfnamefont{G.~F.} \bibnamefont{Sterman}},
  \bibinfo{journal}{Nucl. Phys.} \textbf{\bibinfo{volume}{B261}},
  \bibinfo{pages}{104} (\bibinfo{year}{1985}).

\bibitem[{\citenamefont{Sterman}(1987)}]{Sterman1987}
\bibinfo{author}{\bibfnamefont{G.~F.} \bibnamefont{Sterman}},
  \bibinfo{journal}{Nucl. Phys.} \textbf{\bibinfo{volume}{B281}},
  \bibinfo{pages}{310} (\bibinfo{year}{1987}).

\bibitem[{\citenamefont{Frixione et~al.}(1993)\citenamefont{Frixione, Mangano,
  Nason, and Ridolfi}}]{Frixione1993a}
\bibinfo{author}{\bibfnamefont{S.}~\bibnamefont{Frixione}},
  \bibinfo{author}{\bibfnamefont{M.~L.} \bibnamefont{Mangano}},
  \bibinfo{author}{\bibfnamefont{P.}~\bibnamefont{Nason}}, \bibnamefont{and}
  \bibinfo{author}{\bibfnamefont{G.}~\bibnamefont{Ridolfi}},
  \bibinfo{journal}{Phys. Lett.} \textbf{\bibinfo{volume}{B319}},
  \bibinfo{pages}{339} (\bibinfo{year}{1993}), \eprint{hep-ph/9310350}.

\bibitem[{\citenamefont{de~Florian and Frixione}(1999)}]{Florian1999a}
\bibinfo{author}{\bibfnamefont{D.}~\bibnamefont{de~Florian}} \bibnamefont{and}
  \bibinfo{author}{\bibfnamefont{S.}~\bibnamefont{Frixione}},
  \bibinfo{journal}{Phys.Lett.} \textbf{\bibinfo{volume}{B457}},
  \bibinfo{pages}{236} (\bibinfo{year}{1999}), \eprint{hep-ph/9904320}.

\bibitem[{\citenamefont{Schuler and Sjostrand}(1996)}]{Schuler1996}
\bibinfo{author}{\bibfnamefont{G.~A.} \bibnamefont{Schuler}} \bibnamefont{and}
  \bibinfo{author}{\bibfnamefont{T.}~\bibnamefont{Sjostrand}},
  \bibinfo{journal}{Phys. Lett.} \textbf{\bibinfo{volume}{B376}},
  \bibinfo{pages}{193} (\bibinfo{year}{1996}), \eprint{hep-ph/9601282}.

\bibitem[{\citenamefont{Gl\"{u}ck et~al.}(1999)\citenamefont{Gl\"{u}ck, Reya,
  and Schienbein}}]{Gluck1999}
\bibinfo{author}{\bibfnamefont{M.}~\bibnamefont{Gl\"{u}ck}},
  \bibinfo{author}{\bibfnamefont{E.}~\bibnamefont{Reya}}, \bibnamefont{and}
  \bibinfo{author}{\bibfnamefont{I.}~\bibnamefont{Schienbein}},
  \bibinfo{journal}{Phys. Rev.} \textbf{\bibinfo{volume}{D60}},
  \bibinfo{pages}{054019} (\bibinfo{year}{1999}), \eprint{hep-ph/9903337}.

\bibitem[{\citenamefont{Aversa et~al.}(1989)\citenamefont{Aversa, Chiappetta,
  Greco, and Guillet}}]{Aversa1989}
\bibinfo{author}{\bibfnamefont{F.}~\bibnamefont{Aversa}},
  \bibinfo{author}{\bibfnamefont{P.}~\bibnamefont{Chiappetta}},
  \bibinfo{author}{\bibfnamefont{M.}~\bibnamefont{Greco}}, \bibnamefont{and}
  \bibinfo{author}{\bibfnamefont{J.~P.} \bibnamefont{Guillet}},
  \bibinfo{journal}{Nucl. Phys.} \textbf{\bibinfo{volume}{B327}},
  \bibinfo{pages}{105} (\bibinfo{year}{1989}).

\bibitem[{\citenamefont{J\"{a}ger
  et~al.}(2003{\natexlab{b}})\citenamefont{J\"{a}ger, Sch\"{a}fer, Stratmann,
  and Vogelsang}}]{Jager2003}
\bibinfo{author}{\bibfnamefont{B.}~\bibnamefont{J\"{a}ger}},
  \bibinfo{author}{\bibfnamefont{A.}~\bibnamefont{Sch\"{a}fer}},
  \bibinfo{author}{\bibfnamefont{M.}~\bibnamefont{Stratmann}},
  \bibnamefont{and}
  \bibinfo{author}{\bibfnamefont{W.}~\bibnamefont{Vogelsang}},
  \bibinfo{journal}{Phys. Rev.} \textbf{\bibinfo{volume}{D67}},
  \bibinfo{pages}{054005} (\bibinfo{year}{2003}{\natexlab{b}}),
  \eprint{hep-ph/0211007}.

\bibitem[{\citenamefont{Kidonakis and Sterman}(1997)}]{Kidonakis1997}
\bibinfo{author}{\bibfnamefont{N.}~\bibnamefont{Kidonakis}} \bibnamefont{and}
  \bibinfo{author}{\bibfnamefont{G.~F.} \bibnamefont{Sterman}},
  \bibinfo{journal}{Nucl.Phys.} \textbf{\bibinfo{volume}{B505}},
  \bibinfo{pages}{321} (\bibinfo{year}{1997}), \eprint{hep-ph/9705234}.

\bibitem[{\citenamefont{Kidonakis
  et~al.}(1998{\natexlab{a}})\citenamefont{Kidonakis, Oderda, and
  Sterman}}]{Kidonakis1998a}
\bibinfo{author}{\bibfnamefont{N.}~\bibnamefont{Kidonakis}},
  \bibinfo{author}{\bibfnamefont{G.}~\bibnamefont{Oderda}}, \bibnamefont{and}
  \bibinfo{author}{\bibfnamefont{G.~F.} \bibnamefont{Sterman}},
  \bibinfo{journal}{Nucl. Phys.} \textbf{\bibinfo{volume}{B525}},
  \bibinfo{pages}{299} (\bibinfo{year}{1998}{\natexlab{a}}),
  \eprint{hep-ph/9801268}.

\bibitem[{\citenamefont{Kidonakis
  et~al.}(1998{\natexlab{b}})\citenamefont{Kidonakis, Oderda, and
  Sterman}}]{Kidonakis1998}
\bibinfo{author}{\bibfnamefont{N.}~\bibnamefont{Kidonakis}},
  \bibinfo{author}{\bibfnamefont{G.}~\bibnamefont{Oderda}}, \bibnamefont{and}
  \bibinfo{author}{\bibfnamefont{G.~F.} \bibnamefont{Sterman}},
  \bibinfo{journal}{Nucl. Phys.} \textbf{\bibinfo{volume}{B531}},
  \bibinfo{pages}{365} (\bibinfo{year}{1998}{\natexlab{b}}),
  \eprint{hep-ph/9803241}.

\bibitem[{\citenamefont{Bonciani et~al.}(2003)\citenamefont{Bonciani, Catani,
  Mangano, and Nason}}]{Bonciani:2003nt}
\bibinfo{author}{\bibfnamefont{R.}~\bibnamefont{Bonciani}},
  \bibinfo{author}{\bibfnamefont{S.}~\bibnamefont{Catani}},
  \bibinfo{author}{\bibfnamefont{M.~L.} \bibnamefont{Mangano}},
  \bibnamefont{and} \bibinfo{author}{\bibfnamefont{P.}~\bibnamefont{Nason}},
  \bibinfo{journal}{Phys.Lett.} \textbf{\bibinfo{volume}{B575}},
  \bibinfo{pages}{268} (\bibinfo{year}{2003}), \eprint{hep-ph/0307035}.

\bibitem[{\citenamefont{Sterman and Vogelsang}(2001)}]{Sterman2001}
\bibinfo{author}{\bibfnamefont{G.~F.} \bibnamefont{Sterman}} \bibnamefont{and}
  \bibinfo{author}{\bibfnamefont{W.}~\bibnamefont{Vogelsang}},
  \bibinfo{journal}{JHEP} \textbf{\bibinfo{volume}{0102}}, \bibinfo{pages}{016}
  (\bibinfo{year}{2001}), \eprint{hep-ph/0011289}.

\bibitem[{\citenamefont{Laenen et~al.}(1998)\citenamefont{Laenen, Oderda, and
  Sterman}}]{Laenen1998}
\bibinfo{author}{\bibfnamefont{E.}~\bibnamefont{Laenen}},
  \bibinfo{author}{\bibfnamefont{G.}~\bibnamefont{Oderda}}, \bibnamefont{and}
  \bibinfo{author}{\bibfnamefont{G.~F.} \bibnamefont{Sterman}},
  \bibinfo{journal}{Phys.Lett.} \textbf{\bibinfo{volume}{B438}},
  \bibinfo{pages}{173} (\bibinfo{year}{1998}), \eprint{hep-ph/9806467}.

\bibitem[{\citenamefont{Kidonakis and Owens}(2001)}]{Kidonakis2001}
\bibinfo{author}{\bibfnamefont{N.}~\bibnamefont{Kidonakis}} \bibnamefont{and}
  \bibinfo{author}{\bibfnamefont{J.~F.} \bibnamefont{Owens}},
  \bibinfo{journal}{Phys. Rev.} \textbf{\bibinfo{volume}{D63}},
  \bibinfo{pages}{054019} (\bibinfo{year}{2001}), \eprint{hep-ph/0007268}.

\bibitem[{\citenamefont{Kelley and Schwartz}(2011)}]{Kelley:2010fn}
\bibinfo{author}{\bibfnamefont{R.}~\bibnamefont{Kelley}} \bibnamefont{and}
  \bibinfo{author}{\bibfnamefont{M.~D.} \bibnamefont{Schwartz}},
  \bibinfo{journal}{Phys.Rev.} \textbf{\bibinfo{volume}{D83}},
  \bibinfo{pages}{045022} (\bibinfo{year}{2011}), \eprint{1008.2759}.

\bibitem[{\citenamefont{Catani et~al.}(2013)\citenamefont{Catani, Grazzini, and
  Torre}}]{Catani:2013vaa}
\bibinfo{author}{\bibfnamefont{S.}~\bibnamefont{Catani}},
  \bibinfo{author}{\bibfnamefont{M.}~\bibnamefont{Grazzini}}, \bibnamefont{and}
  \bibinfo{author}{\bibfnamefont{A.}~\bibnamefont{Torre}}
  (\bibinfo{year}{2013}), \eprint{1305.3870}.

\bibitem[{\citenamefont{Catani et~al.}(1996)\citenamefont{Catani, Mangano,
  Nason, and Trentadue}}]{Catani1996a}
\bibinfo{author}{\bibfnamefont{S.}~\bibnamefont{Catani}},
  \bibinfo{author}{\bibfnamefont{M.~L.} \bibnamefont{Mangano}},
  \bibinfo{author}{\bibfnamefont{P.}~\bibnamefont{Nason}}, \bibnamefont{and}
  \bibinfo{author}{\bibfnamefont{L.}~\bibnamefont{Trentadue}},
  \bibinfo{journal}{Nucl. Phys.} \textbf{\bibinfo{volume}{B478}},
  \bibinfo{pages}{273} (\bibinfo{year}{1996}), \eprint{hep-ph/9604351}.

\bibitem[{\citenamefont{Catani et~al.}(1998)\citenamefont{Catani, Mangano, and
  Nason}}]{Catani1998a}
\bibinfo{author}{\bibfnamefont{S.}~\bibnamefont{Catani}},
  \bibinfo{author}{\bibfnamefont{M.~L.} \bibnamefont{Mangano}},
  \bibnamefont{and} \bibinfo{author}{\bibfnamefont{P.}~\bibnamefont{Nason}},
  \bibinfo{journal}{JHEP} \textbf{\bibinfo{volume}{9807}}, \bibinfo{pages}{024}
  (\bibinfo{year}{1998}), \eprint{hep-ph/9806484}.

\bibitem[{\citenamefont{Kidonakis and Owens}(2000)}]{Kidonakis2000}
\bibinfo{author}{\bibfnamefont{N.}~\bibnamefont{Kidonakis}} \bibnamefont{and}
  \bibinfo{author}{\bibfnamefont{J.}~\bibnamefont{Owens}},
  \bibinfo{journal}{Phys.Rev.} \textbf{\bibinfo{volume}{D61}},
  \bibinfo{pages}{094004} (\bibinfo{year}{2000}), \eprint{hep-ph/9912388}.

\bibitem[{\citenamefont{Catani et~al.}(1999)\citenamefont{Catani, Mangano,
  Nason, Oleari, and Vogelsang}}]{Catani1999}
\bibinfo{author}{\bibfnamefont{S.}~\bibnamefont{Catani}},
  \bibinfo{author}{\bibfnamefont{M.~L.} \bibnamefont{Mangano}},
  \bibinfo{author}{\bibfnamefont{P.}~\bibnamefont{Nason}},
  \bibinfo{author}{\bibfnamefont{C.}~\bibnamefont{Oleari}}, \bibnamefont{and}
  \bibinfo{author}{\bibfnamefont{W.}~\bibnamefont{Vogelsang}},
  \bibinfo{journal}{JHEP} \textbf{\bibinfo{volume}{9903}}, \bibinfo{pages}{025}
  (\bibinfo{year}{1999}), \eprint{hep-ph/9903436}.

\bibitem[{\citenamefont{Tung et~al.}(2007)\citenamefont{Tung, Lai, Belyaev,
  Pumplin, Stump et~al.}}]{Tung2007}
\bibinfo{author}{\bibfnamefont{W.}~\bibnamefont{Tung}},
  \bibinfo{author}{\bibfnamefont{H.}~\bibnamefont{Lai}},
  \bibinfo{author}{\bibfnamefont{A.}~\bibnamefont{Belyaev}},
  \bibinfo{author}{\bibfnamefont{J.}~\bibnamefont{Pumplin}},
  \bibinfo{author}{\bibfnamefont{D.}~\bibnamefont{Stump}},
  \bibnamefont{et~al.}, \bibinfo{journal}{JHEP}
  \textbf{\bibinfo{volume}{0702}}, \bibinfo{pages}{053} (\bibinfo{year}{2007}),
  \eprint{hep-ph/0611254}.

\bibitem[{\citenamefont{de~Florian et~al.}(2007)\citenamefont{de~Florian,
  Sassot, and Stratmann}}]{Florian2007}
\bibinfo{author}{\bibfnamefont{D.}~\bibnamefont{de~Florian}},
  \bibinfo{author}{\bibfnamefont{R.}~\bibnamefont{Sassot}}, \bibnamefont{and}
  \bibinfo{author}{\bibfnamefont{M.}~\bibnamefont{Stratmann}},
  \bibinfo{journal}{Phys.Rev.} \textbf{\bibinfo{volume}{D75}},
  \bibinfo{pages}{114010} (\bibinfo{year}{2007}), \eprint{hep-ph/0703242}.

\end{thebibliography}

\end{document}